\documentclass{aa}
\usepackage{subfig}
\usepackage{graphicx}
\usepackage{txfonts}
\usepackage{enumerate}
\usepackage{color}

\graphicspath{ {images/} }
\usepackage{natbib,twoopt}
\usepackage[breaklinks=true]{hyperref} 
\bibpunct{(}{)}{;}{a}{}{,} 
\makeatletter
\newcommandtwoopt{\citeads}[3][][]{\href{http://adsabs.harvard.edu/abs/#3}%
{\def\hyper@linkstart##1##2{}%
\let\hyper@linkend\@empty\citealp[#1][#2]{#3}}}
\newcommandtwoopt{\citepads}[3][][]{\href{http://adsabs.harvard.edu/abs/#3}%
{\def\hyper@linkstart##1##2{}%
\let\hyper@linkend\@empty\citep[#1][#2]{#3}}}
\newcommandtwoopt{\citetads}[3][][]{\href{http://adsabs.harvard.edu/abs/#3}%
{\def\hyper@linkstart##1##2{}%
\let\hyper@linkend\@empty\citet[#1][#2]{#3}}}
\newcommandtwoopt{\citeyearads}[3][][]%
{\href{http://adsabs.harvard.edu/abs/#3}
{\def\hyper@linkstart##1##2{}
\let\hyper@linkend\@empty\citeyear[#1][#2]{#3}}}
\makeatother

\begin{document}
 \include{commands}
\titlerunning{MORESANE}
\title{MORESANE: MOdel REconstruction by Synthesis-ANalysis Estimators}
\subtitle{A sparse deconvolution algorithm for radio interferometric imaging}

\author{A.~Dabbech\inst{1}\thanks{e-mail: arwa.dabbech@oca.eu}
        \and C.~Ferrari\inst{1}
        \and D.~Mary\inst{1}
        \and E.~Slezak\inst{1}
        \and O.~Smirnov\inst{2,3}
        \and {J.S.~Kenyon\inst{2}}
        }
\institute{Laboratoire Lagrange, UMR7293, Universit\'e Nice Sophia-Antipolis, CNRS, Observatoire de la C\^ote d'Azur, F-06300 Nice, France \and Centre for Radio Astronomy Techniques \& Technologies (RATT), Department of Physics and Electronics, Rhodes University, PO Box 94, Grahamstown 6140, South Africa 
\and SKA South Africa, 3rd Floor, The Park, Park Road, Pinelands, 7405, South Africa
} 

\date{Received July, 2014}

\abstract{{  The current years are seeing huge developments of radio telescopes and a tremendous increase of their capabilities (sensitivity, angular and spectral resolution, field of view, \dots). Such systems make mandatory the design of more   sophisticated techniques not only  for transporting, storing and processing this new generation of radio interferometric data, but also for restoring the astrophysical information contained in such data.}}{{ In this paper we present a new radio deconvolution algorithm named MORESANE { and its application to fully realistic simulated data of MeerKAT, one of the SKA precursors}. This method has been designed for the difficult case of restoring  diffuse astronomical sources which are faint in brightness, complex in morphology and possibly buried in the dirty beam's side lobes of bright radio sources in the field.}}{{ MORESANE {is a greedy algorithm which} combines complementary types of sparse recovery methods in order to reconstruct the most appropriate sky model from observed radio visibilities. A synthesis approach is used for the reconstruction of images, in which the  synthesis atoms representing the unknown sources are learned using analysis priors. We apply this new deconvolution method to fully realistic simulations of radio observations of a galaxy cluster and of an HII region in M31.}}{{ We show that MORESANE is able to efficiently reconstruct images composed from a wide variety of sources (compact point-like objects, extended tailed radio galaxies, low-surface brightness emission) from radio interferometric data.  Comparisons with other available algorithms, which include  multi-scale CLEAN and the recently proposed methods by Li et al. (2011) and Carrillo et al. (2102), indicate that MORESANE provides competitive results in terms of both total flux/surface brightness conservation and fidelity of the reconstructed model. MORESANE seems particularly well suited for the recovery of diffuse and extended sources, as well as bright and compact radio sources known to be hosted in galaxy clusters.}}{}
\maketitle
\section{Introduction}
In the last 40 years, the radio community has mainly been using as a reliable and well understood method for deconvolving interferometric data the CLEAN algorithm, and its different (including multi-resolution) variants \citep[e.g.][]{Hogbom74,Wakker88,Cornwell08}. Even if other methods have been designed during this period \citep[see for instance][]{Magain98,Pirzkal00,Starck02,Giovannelli05} none has become  in practice as popular and as widely used as CLEAN. 

Deep and/or all--sky radio surveys characterized by sub--mJy sensitivity and arcsec angular resolution, as well as by high ($>$1000) signal--to--noise and spatial dynamic ranges (challenging features for a proper deconvolution and reconstruction of both bright and diffuse radio components) will be available in the next decades thanks to incoming and future radio facilities, such as the Low Frequency Array (LOFAR), the Australian Square Kilometre Array Pathfinder (ASKAP, Australia) and the Karoo Array Telescope (MeerKAT, South Africa). These revolutionary radio-telescopes, operating in a wide region of the electromagnetic spectrum (from 10 MHz to 15 GHz), are the technical and scientific pathfinders of the Square Kilometre Array (SKA). With a  total collecting area of one square kilometre, SKA will be the largest telescope ever built (see 
\url{https://www.skatelescope.org/}).

Recently, great attention has been paid in various fields of the signal/image processing community to the theory of Compressed Sensing  \citep[CS,][]{donoho2006compressed,candes2006robust}. Although the current theoretical results of CS do not provide means for reconstructing more accurately realistic radio interferometric images, there is one key domain which allows to do so, and the ``applied CS'' literature often builds on this domain: sparse representations. Sparse approximation through dedicated representations is a theory {\it{per se}} and has a very long history \citep{Mallat08}. In the signal processing community, and in particular in denoising and compression, sparsity principles opened a new era when Donoho and Johnstone \citep{donoho93} proved the minimax optimality of thresholding rules in wavelet representation spaces. We propose in Sect.\,\ref{sparse} a survey of the use of sparse representations in radio interferometry.

In this paper we describe a new deconvolution algorithm named ``MORESANE'' (MOdel REconstruction by Synthesis-ANalysis Estimators''), which combines complementary types of sparse representation priors. MORESANE has been designed for the restoration of faint diffuse astronomical sources, with a particular interest to recover the diffuse intracluster radio emission from galaxy clusters. These structures are known to host a variety of radio sources, such as : compact and bright radio galaxies, which can present a tailed morphology modeled by the interaction with the intra-cluster medium \citep[ICM; e.g.][]{2002ASSL..272..163F}; radio bubbles filling holes in the ICM distribution and rising buoyantly through the thermal gas observed in X--rays \cite[e.g.][]{2012A&A...547A..56D}; Mpc-scale very low surface brightness sources of radio emission, related to the presence of relativistic electrons (Lorentz factors $\gamma \gg 1000$) and weak magnetic fields ($\sim \mu$Gauss) in the ICM \citep[e.g.][]{2008SSRv..134...93F}. 

An increased capability of the detection of diffuse radio emission is of course of great relevance not only for galaxy cluster studies, but also for other astronomical research areas, such as supernova remnants \citep[e.g.][]{2014MNRAS.440.3220B}, radio continuum emission from  the Milky Way \citep[e.g.][]{1985IAUS..106..239B}, star forming regions in nearby galaxies \citep[e.g.][]{2006A&A...456..847P} and possibly, in the next future, filaments of diffuse radio emission related to electron acceleration in the cosmic web (Vazza et al. 2014, in prep.).

After an introduction about the radio interferometric model and sparse representations, in Sects.\,\ref{radiop} and \ref{sparse} respectively, we motivate and describe our new algorithm in Sects.\,\ref{motiv} and \ref{moresane} respectively. Applications of MORESANE {\color{black}to both simplified }and fully realistic simulations of test images are presented in Sect.\,\ref{simulations}. We conclude with a discussion of our results and we list several evolutions for MORESANE (Sect.\,\ref{Conclusions}).

A word on the notations before starting. We shall denote matrices by bold upper case letters (e.g., ${\bf{M}}$), vectors either by bold lower case letters (e.g., ${\bf{v}}$) or by indexed matrix symbols when they correspond to a column of a matrix (e.g., ${\bf{M}}_i$ is the $i^{\textrm{th}}$ column of  ${\bf{M}}$). Scalars (and complex numbers) are not bolded except if they correspond to components of a vector (e.g., ${\bf{v}}_i$  is the $i^{\textrm{th}}$ component of ${\bf{v}}$) or of a matrix (e.g., ${\bf{M}}_{ij}$  is the component at the $i^{\textrm{th}}$ row and $j^{\textrm{th}}$ column of ${\bf{M}}$).

\section{Radio interferometric imaging} 
\label{radiop}
Radio interferometric data are obtained from the response of the radio interferometer to the electric field  coming from astrophysical sources. The electro-magnetic radiation emitted by all the observed celestial sources will arrive at an observation point $\bf r$ producing a total received electric field ($E_{\nu}({\bf r})$) that we will consider as a scalar and quasi monochromatic quantity. For the sake of simplicity, in the following we will omit the index $\nu$.

For an interferometer, each radio measurement, called complex visibility, corresponds to the spatial coherence of the electric field  measured by a pair of antennas which have  coordinates $\bf r_1$ and $\bf r_2$ \citep{Thompson}.
\begin{equation}
 V({\bf r_1},{\bf r_2})=\left\langle { E}({\bf r_1}){ E}^*({\bf r_2}) \right\rangle ,
\end{equation}
where $\left\langle \cdot \right\rangle $ represents time averaging and $^*$ complex conjugate.

The spatial coherence function of the electric field  ${ E}$ depends only on the  baseline vector ${\bf r_1}-{\bf r_2}$ and it is correlated to the intensity distribution of incoming radiation $I$({\textbf{s}}) (where \textbf{s} is the unit vector denoting the direction on the sky) through:

\begin{equation}
\label{visth}
 V({\bf r_1},{\bf r_2})\approx \int I(\text{\textbf{s}}) e^{-2\pi i \nu \textbf{s}^\top({\bf r_1-r_2})/c} d\Omega.
  \end{equation}
In the equation above  $^\top$ stands for transpose, $c$ is the speed of light, $d\Omega$ is the differential solid angle and we assume an isotropic antenna response.
Since interferometer antennas have a finite size, an additional factor can enter in \eqref{visth}, that is the primary beam pattern, which describes the sensitivity of the interferometer elements as a function of direction \textbf{s}.

In the previous equation, the baseline vector ${\bf b}={\bf r_1}-{\bf r_2}$ can be expressed with components measured in units of wavelength $(u,v,w)$, where $w$ points in the direction of the line of sight and $(u,v)$ lie on its perpendicular plane. The direction cosines $(l,m,n)$ define the  position of a distant source on the celestial sphere, with $(l,m)$ measured with respect to $(u,v)$ axis. Note that in the adopted formalism $l^2+m^2+n^2=1$ and so the coordinates $(l,m)$ are sufficient to specify a given point in the celestial sphere. 
Using this formalism, \eqref{visth} can be written as:
\begin{equation}
\label{vis}
 V(u,v,w)= \int \int I(l,m)~ e^{-2\pi i (ul+vm+wn)} \frac{dl~dm}{\sqrt{1-l^2-m^2}}.
  \end{equation}
In the particular cases where all measurements are acquired in a plane (i.e. $w=0$, such as with East-West interferometers) and/or the sources are limited to a small region of the sky (i.e. $n \simeq 1$, for small fields-of-view that is the case considered in this paper), \eqref{vis} reduces to a two-dimensional Fourier transform.

In the impossible case of visibilities measured on the whole $(u,v)$ plane, inverse Fourier transform of $V(u,v)$ would thus directly yield to the sky brightness image $I(l,m)$. In practice, visibilities are measured at particular points of  the Fourier  domain defining the $(u,v)$ {coverage} of the observations.  The set of samples depends on the configuration and number of the antennas, the time grid of measurements and the number of channels, as the baselines change with the Earth's rotation. A sampling function  $M(u,v)$ is thus introduced, which is  composed of Dirac delta function where visibilities are acquired. 

After the necessary  calibration step on the visibilities \citep[that is not described here; see e.g.][]{Fomalont99}, the measured visibilities can be written as:
\begin{equation}
\label{Vis_real}
{ V}_{mes}={ M}\cdot ({ V} + {{\epsilon}}),
 \end{equation} 
 where $\epsilon$ corresponds to a white Gaussian noise coming essentially from the sky, receivers and ground pick up. Note in addition, that a weighted sampling function can be applied to the data, with different weights assigned to different observed visibilities depending on their reliability,  their $(u,v)$ locus (tapering function) or  their density in the $(u,v)$ plane (density weighting function), \citep{Briggs99}.

The image formed by taking the inverse Fourier transform of $V_{mes}$ is called  \textit{dirty image}, which is defined as the convolution of the true sky surface brightness distribution ${I}(l,m) $ with  the Fourier inverse  transform of the sampling function $M(u,v)$ (known as the \textit{dirty beam} or the Point Spread Function (PSF) of the array). In practice, Fast Fourier Transforms (FFT) are used in which observed visibilities require to be interpolated on a regular grid of $2^N \times 2^M$ points, generating a $N \times M$ pixel image with a pixel size taken to be smaller ($\sim 1/3 ~-~ 1/5$) as compared to the angular resolution of the instrument. Different ways can be adopted to optimize the FFT interpolation \citep{Briggs99}, whose discussion goes beyond the purpose of this paper.

In this framework, the model for the visibility measurements can be written in matrix form as:
\begin{equation}
\label{classical_model0}
{\bf {v}}= {\bf {MFx}}+{\bf M}\pmb{\epsilon},
\end{equation}
where ${\bf {v}} \in\mathbb{R}^N$ is a column vector which contains the measured visibilities for the sampled frequencies and zeros otherwise.  $\bf M$  is a diagonal matrix with $0$ and $1$ on the diagonal, which expresses  the incomplete sampling of the spatial frequencies.  $\bf F$ (resp. $\bf F^{\dagger}$) corresponds to the Fourier (resp. Fourier inverse) transform, and the vector ${\bf x} \in\mathbb{R}^N $ is the  sky brightness image. Equivalently, the dirty image $\bf y$ is obtained by inverse Fourier transform of the sparse visibility map:
\begin{equation}
\label{classical_model}
{\bf {y}}={\bf{F^{\dagger}}} {\bf {v}}= {\bf {Hx+n}},
\end{equation}
where ${\bf H}={\bf{F^{\dagger}MF}},  {\bf H}\in{\mathbb{R}}^N\times{\mathbb{R}}^N $ is the convolution operator corresponding to the array's PSF and ${\bf n}\in\mathbb{R}^N$ is the noise in the image domain.
 In this setting $\bf {H}$ is a circulant matrix operator, of which every column is a shifted version of the PSF for every pixel position. 
 Note finally that in model \eqref{classical_model}, the noise is additive Gaussian and correlated  because of  the missing points in the $(u,v)$ domain \citep{Thompson}.  

We will hereafter  restrict ourselves to the simplified acquisition model described above. As we shall see, accurate image deconvolution is already challenging in this case, especially for astrophysical scenes containing faint diffuse sources together with brighter and more compact ones. 
 \section{Sparse representations in radio interferometry }
 \label{sparse}
 \subsection{CLEAN}
 \label{CLEANs}
 Radio interferometry has a long term acquaintance with sparse representations.   H\"ogbom's CLEAN algorithm \citep{Hogbom74} and the family
 of related methods \citep{Wakker88,Bijaoui92,cornwell,schwarz} implement ideas similar to Matching Pursuit \citep{friedmanmp,mallat93} and to $\ell_1$ penalization \citep{solo}. In fact, ``CLEAN'' refers in the radioastronomical community to a family of algorithms (Clark's CLEAN, Cotton-Schwab CLEAN, MultiResolution CLEAN, etc...).

A remarkable fact is that the CLEAN method remains a reference and  a very well known tool for almost all radio astronomers. There may be several reasons for this fact.
First, CLEAN is a competitive algorithm, with best results on point like sources and less accurate recovery of extended sources. In CLEAN, the ``CLEAN factor'' does a lot:  following H\"ogbom's original version of the algorithm, the point source's contribution which is the most correlated to the data is only partly subtracted from the data (in contrast  to Matching Pursuit, which makes the residual orthogonal to this atom).
This has the effect to create ``detections'' at many locations and to mitigate the influence of the brightest sources. These localized numerous spikes mimic extended flux components and somewhat compensate the point-like synthesis of the restored image once the detection is reconvolved by the clean beam. Besides, after the stopping criterion is met, the residual is added to the restored image, with the same compensating effect. \\
From a practical viewpoint, CLEAN is easy to implement and does not require any optimization knowledge. It is also easy to build modular versions of CLEAN with deconvolution by patches for instance, allowing to account for direction dependent effects and to couple image restoration and calibration processes \citep{tasse2013}. 

Finally, the greedy structure of CLEAN was probably a major advantage for devising an operational spatio-spectral radio deconvolution algorithm (to our knowledge the only algorithm allowing to deconvolve visibility data cubes), the Multiscale-Multifrequency CLEAN implemented in LOFAR data processing \citep{MPvanHaarlem:2013gi}. As a matter of fact, CLEAN algorithms are implemented in many standard radio imaging softwares.
\subsection{Recent works: sparse representations}
\label{recent}
In the second half of the 2000's, stellar interferometry has been identified as a typical instance of Compressed Sensing \citep[CS,][]{candes2006robust, donoho2006compressed} acquisition.
Since the  theoretical results of CS had shed a new mathematical light on random Fourier sampling  of sparse spikes, radio interferometry has
 appeared as a natural case of CS and major achievements were foreseen in this domain from CS theory. 
 Looking back at the literature from this period to now, it seems that  innovation in recent radio interferometric reconstruction  methods has less grown from CS theorems (because their assumptions are most often not satisfied in practical situations) than from 
an unchained research activity in sparse representations and convex optimization \citep{SPARCS13}. Although these domains existed long before CS, they certainly benefited from the CS success. A survey of the evolution of sparse models in the recent literature of radio interferometric image reconstruction is proposed below. These models fall in two categories, sparse analysis or sparse synthesis, a vocabulary which originate from frame theory and was studied in the context of sparse representations by \citet{Elad06}.
\subsection{Sparse synthesis }
This approach assumes that the image to be restored, $\bf x$, can be sparsely synthesized by few elementary features called \textit{atoms}. More precisely, ${\bf x}$ is assumed to be a linear combination of a few columns of some full rank $\ell_2$- normalized dictionary  $\bf S$, of size $(N,L)$, with  $L$ usually greater than  $N$ :  
\begin{equation}
{\bf x}={\bf S} \pmb{\gamma}, \textrm{ \quad where \;} {{\pmb{\gamma}}} \in \mathbb{R}^L \textrm{ \; is sparse}.
\label{sp}
\end{equation}
With \eqref{sp},  model \eqref{classical_model0}
becomes:
\begin{equation}
\label{model-S}
\bf v= {\bf{MFS}}\pmb{\gamma} +\pmb{\epsilon}, ~~{\displaystyle\rm with}~~ \pmb{\gamma} ~\displaystyle\rm {sparse.}
\end{equation} 
The most simple and intuitive sparsity measure is the number of non-zero entries of ${\bf x}$ (i.e., the $\ell_0$ pseudo-norm), but  $\ell_0$ is not convex. To benefit from the properties of convex optimization, the $\ell_0$ penalization is often 
relaxed and replaced by $\ell_1$\footnote{For a vector $\bf{x}$, $\ell_p^p=\sum_i | \bf{x}_i |^p$.}, which still promotes strict sparsity and thus acts as a variable selection procedure ($\ell_p^p$ with  $0<p<1$ also, but leads to more difficult -non convex- optimization problems; $\ell_p^p,~p>1$ does not).\\ 

In a sparsity-regularized reconstruction approach, a typical regularization term corresponding to such penalties (but there are many others) is of the form $\mu_p\parallel \pmb{\gamma}\parallel_p^p$, with a regularization parameter $\mu_p \in \mathbb{R}_+$ and it is  added to the data fidelity term (the squared Euclidean norm of the error for i.i.d. Gaussian noise). The  vectors $\pmb{\gamma}$ that will minimize the cost function:
\begin{equation}
\label{JMAP-S'}
{j}_{s}({\pmb \gamma})=\frac{1}{2}\|{\bf {\bf{MFS}}\pmb{\gamma}-v}\|_2^2+\mu_p \parallel{\bf \pmb{\gamma}}\parallel_p^p\;, \quad 0\leq p \leq 1,
\end{equation}
will then tend to be  sparse for sufficiently large values of $\mu_p$.
Synthesis-based approaches lead  thus to solutions of the form\footnote{${\bf{MFS}}$ is assumed to have unit-norm $\ell_2$ columns. If this is not the case,  the components of ${\pmb{\gamma}}$ should be weighted accordingly  \citep[see e.g.][]{2011ISTSP}.}
\begin{equation}
\label{MAP-S'}
{\bf x}_{s}^*= {\bf S}\cdot\lbrace{\displaystyle\rm arg}\displaystyle\rm \min_{\pmb{\gamma}}\frac{1}{2}\|{\bf {\bf{MFS}}\pmb{\gamma}-v}\|_2^2 +\mu_p \parallel{\bf \pmb{\gamma}}\parallel_p^p\rbrace.
\end{equation}
The minimization problem corresponding to the particular case  $p=1$ is called Basis Pursuit DeNoising (BPDB) in optimization  \citep{chen1998atomic}.
\subsection{Sparse Analysis }
In this approach, regularity conditions on ${\bf x}$ are imposed by an operator $ {\bf A}^\top$. The sparse analysis approach consists in finding a solution $\bf x$ that is not correlated to some atoms (columns) of a dictionary $ {\bf A}$ of size $ (N,L)$. Hence, the sparse analysis model assumes  that  $ {\bf A}^\top {\bf x}$ is sparse.

Adopting a regularization term imposing this sparsity constraint, sparse analysis approaches usually seek solutions of the form:
\begin{equation}
\label{A-MAP'}
{\bf x}_{a}^*= {\displaystyle\rm arg}\displaystyle\rm \min_{\bf x}\frac{1}{2}\|{\bf MFx-v}\|_2^2 +\mu_p \parallel{\bf A}^\top {\bf x}\parallel_p^p\; , \quad 0\leq p \leq 1.
\end{equation}
\subsection{Representations and dictionaries}
\label{lesdec}
The sparsity  expressed through $\bf{S}$ on $\pmb{\gamma}$ or on $\bf{A}^\top  {\bf x}$  imposes  that the signal is characterized by low dimensional subspaces. 
They can be orthonormal  transforms (corresponding to orthonormal bases), or more generally redundant (overcomplete) dictionaries.
These subspaces correspond to mathematical representations of the signal : the columns of $\bf{S}$ correspond to geometrical features that are likely to describe the unknown signal or image, while the columns of $\bf{A}$ impose  geometrical constraints (in analysis).

A large variety of such representations has been elaborated in the image processing literature, e.g.,   canonical basis indeed (corresponding to point-like structures), Discrete Cosine   Transform (DCT, 2-D plane waves), wavelets (localized patterns in time and  frequency), isotropic undecimated wavelets \citep{isotropic}, curvelets \citep[elongated and curved patterns][]{curvelets}, ridgelets \citep{ridgelets}, shapelets \citep{shapelets} and others \citep[see ][for details on these representations and their applications.]{Mallat08,starckbook}.

The choice of  a dictionary  is made with respect to a {{class of images}}.  {{In Astronomy, wavelets dictionaries are widely used, but they  are known to fail representing well anisotropic structures. In such cases other transforms can be used, that have been designed to capture main features of specific classes of objects.  Among them, curvelets sparsify well curved, elongated patterns (such as planetary rings or thin galaxy arms), while shapelets sparsify well, for instance, various galaxy morphologies. All of them have shown empirical efficiency and can be used in the dictionary.}}
 
 In order  to accurately model complex images with various features, one possibility is indeed to concatenate  several dictionaries  into a larger dictionary. 
However, the   efficiency of a dictionary also critically depends on its size and on the {{existence of fast operators}},
without which restoration algorithms (that are \textit{iterative}) cannot converge in a reasonable time.
 Concatenation or unions of representation spaces are now classically used in denoising and inverse problems because they allow to better account for more complex morphological features than standard transforms used separately 
 \citep[an approach early advocated in \citet{mallat93} and \citet{chen1998atomic}, see also ][]{donohohuo2001,gribonval2003sparse,starckbook}. Such unions may allow to maintain a reasonable computational cost if fast transforms are associated to each representation space. They also provide a natural feature separation through the decomposition coefficients associated to each sub-dictionary.  This property
 is indeed of particular interest in Astronomy, where a celestial scene may contain features  as different as point-like sources, rings, spirals,  or smooth and diffuse components, with various spatial extensions.
\subsection{Synthesis versus Analysis}
\label{avs}
Analysis and synthesis priors lead to different solutions (and algorithms) for redundant dictionaries. When ${\bf{A}}$ and ${\bf{S}}$ are square and invertible, as for instance orthonormal bases, a change of variables with   ${\bf{S}}^{-1}={\bf{A}}^\top$ shows that the approaches in \eqref{MAP-S'} and \eqref{A-MAP'} are equivalent. 
A seminal study was proposed in \citet{Elad06}, whose first result shows that when $\bf S$ is taken as ${{\bf A}^\top}^\dagger$ (the pseudo-inverse of ${\bf{A}}^\top$), the analysis model is restricted to a space of lower dimension than the synthesis one. 
More generally, Theorem 4 of the same paper shows by more involved means that, for $p=1$ and $L\geq N$, there exists  for any $\ell_1$ MAP-analysis problem with full-rank analyzing operator ${\bf A}^\top$ a dictionary ${\bf S}({\bf A}^\top)$ describing an equivalent $\ell_1$ MAP-synthesis problem. The converse is not true. In this sense, sparse synthesis is more general than analysis and allows in theory better reconstruction results.

The question of how the two approaches compare in practice for usual transforms remains however open, even  for the case $p=1$. The works of \citet{carlavan2010} propose an interesting numerical comparison of the two approaches for various transforms and dictionaries in the framework of noisy deconvolution. Their conclusion is that synthesis approaches seem to be preferable at lower noise levels, while analysis is more robust  at higher noise regimes.
 
\citet{candes2010}  report  numerical experiments with redundant dictionaries showing empirically that $\ell_1$-synthesis may perform similarly well as  $\ell_1$-analysis, while other papers highlight better results for analysis models \citep{Carrillo12}. A clear and well identified issue with synthesis is that the number of unknown (synthesis coefficients) may rapidly become prohibitive for large dictionaries, while in analysis the number of unknown remains constant (as it corresponds to the number of image parameters in ${\bf{x}}$). On the other hand, sticking to a synthesis approach with dictionaries of insufficiently many atoms may lead to rough and schematic reconstructed sources. Obtaining  more  theoretical and general results on the analysis vs synthesis comparison  is a very interesting, active and growing subject of research, see for instance the references of \url{http://small-project.eu/publications}.
 
In radio interferometry, each recent reconstruction algorithm has its own sparse representation model. Explicit sparse priors were indeed first expressed   in the direct image space, which is typically appropriate for (but limited to)  fields of unresolved stars \citep{marygretsi2007,maryada2008}. In this case, the restored image can be obtained by solving the BPDN problem associated to \eqref{MAP-S'} with $\bf{S=I}$ (or to \eqref{A-MAP'} with $\bf{A=I}$). This is also the approach of \citet{Wenger10}. 

In order to efficiently recover more complex images, sparse synthesis models involving a dictionary $\bf S$ taken as union of bases with a union of canonical, DCT and orthogonal wavelets bases  were proposed in \citet{aip09,2010SPIE,2010ada1}.  The restored image is in this case obtained by solving (\ref{MAP-S'}) with $p=1$. The ``Compressed Sensing imaging technique BP$^+$'' of \citet{Wiaux09a, Wiaux09b} solves a synthesis problem \eqref{MAP-S'} with $p=1$ subject to an image positivity constraint and $\bf S $ is a redundant dictionary of wavelets.
 
In \citet{Li11}, the ``Compressed Sensing-based deconvolution'' uses for $\bf S$  the Isotropic Undecimated Wavelet Transform \citep[IUWT, ][]{starckbook} and solves \eqref{MAP-S'} under a positivity constraint. We will show results of this method in the simulations.
The works of \citet{McEwen11} consider an analysis-based prior (total variation), for which $\bf{A}^\top$ implements the $\ell_1$ norm of the discrete image gradient:
\begin{equation}
{\bf x}^*_{a}= {\displaystyle\rm arg} \min_{\bf x} \|{\bf x}\|_{TV} \text{\;\;s.t.\;\;} \| {\bf y} - {\bf W \bf M \bf F  \bf x}||_2 < \epsilon,
\end{equation}
where $\epsilon$ is a prescribed fidelity threshold.

Recently, the ``Sparsity Averaging Reweighted Analysis" (SARA)  \citep{Carrillo12} focused on an analysis criterion with a solution of the form  \eqref{A-MAP'} with $p=1$, a positivity constraint on $\bf{x}$ and a union of wavelet bases for $\bf A $. The work of \citet{purify} presents large scale optimization strategies dedicated to this approach. Clearly, sparse models allied to optimization techniques have attracted a lot of attention in this field during  the last 7 or 8 years.
\section{Motivation for an analysis-by-synthesis approach}
\label{motiv}
The imaging system described in Sect.\,\ref{radiop} describes a linear filter whose transfer function is described by the diagonal of $\bf  M$. In the Fourier space, this transfer function has many zeros, making the  problem of reconstructing ${\bf{x}}$ from  ${\bf{y}}$  under-determined and ill-posed. In the image space, the PSF has typically numerous and slowly decreasing sidelobes due to the sparse sampling performed by the interferometer. The PSF extension and irregularity make the recovery of faint objects particularly difficult when  surrounding sources that are orders of magnitude brighter.

The specific problem of restoring faint extended sources  submerged by the contribution of the sidelobes of brighter and more compact sources  has lead us to explore in \citet{dabbechicassp2012} a fast restoration method, which exploits positivity and sparse priors in a hybrid manner, and of which MORESANE is an elaborated version. {{Several essential changes have been introduced in MORESANE w.r.t. the prototype algorithm by \citet{dabbechicassp2012} in order to be able to apply it on realistic radio interferometric data. The main developments include:
\begin{itemize}
\item the identification of the brightest object which done now by taking into account the PSF behaviour in the wavelet domain;
\item the use of parameters $\tau$  and $\gamma$ introduced to improve rapidity and obtain a more accurate estimation of the sky image;
\item the use of the Conjugate gradient instead of the Projected Landweber algorithm for rapidity;
\item the deconvolution scale by scale, which does not oblige the user to specify the number of scales in the IUWT.
\end{itemize}}}

Comprehensive surveys of the vast literature in image reconstruction methods for radio interferometric data can be found in \citet{Starck02} (including methods from model fitting to non parametric deconvolution)
and in \citet{Giovannelli05}, who emphasize  the particular problem of reconstructing complex (both extended and compact) sources. On this specific topic, very few other works can be found. In \citet{Magain98} and \citet{Pirzkal00}, the point-like sources are written in a parametric manner based on amplitudes and peaks positions. Extended morphologies are accounted for in \citet{Magain98} using a Tikhonov regularization, and in \citet{Pirzkal00} by introducing a Gaussian correlation. In \citet{Giovannelli05}, a global criterion is minimized (subject to a positivity constraint), where the penalization term is the sum of the  $\ell_1$ norm of the point-like component and the $\ell_2$ norm of the gradient of the extended component. 

Radio interferometric image reconstruction is a research field where  synthesis and analysis sparse representations
have been extensively and (in fact almost exclusively) investigated in the last few years.
To be efficient on recovering {faint, extended and irregular} sources in high dynamic range scenes, the approach we propose is hybrid in its sparsity priors and somewhat builds on ideas of \citet{Hogbom74} and \citet{stomp}. We use a synthesis approach for the reconstruction of the image but we do not assume that synthesis atoms describing the sources are  fixed in advance, in order to allow more flexibility in the source modelling. The synthesis dictionary atoms are  learned iteratively using analysis-based priors. {{This iterative approach is greedy in nature and thus does not rely on global optimization procedures (such as l1 analysis or synthesis minimization) of any kind}}. The iterative process is important to cope with high dynamic scenes. The analysis approach allows a fast reconstruction.

\section{Model Reconstruction by Synthesis-Analysis Estimators} \label{moresane}
We  model the reference scene $\bf x$ as the superposition  of $ P $ objects:
\begin{equation}
\label{model}
{\bf x}=\sum_{t=1}^P {\pmb \theta}_t{\bf X}_t={\bf X} ~{\pmb \theta},
\end{equation}
where ${\bf X}_t$ , columns of $\bf X$, are $\ell_2$-normalized objects composing $\bf x$. An object may be a single source, or a set of sources sharing similar characteristics in terms of spatial extension and brightness.  The matrix $\bf X$ is an unknown synthesis dictionary of size ($N, P$), $P << N $ and $\pmb \theta $ is a vector of amplitudes of size $P$ with entries ${\pmb{\theta}}_t$.
\noindent
The radio interferometric model \eqref{classical_model} becomes:
\begin{equation}
\label{synth_pb}
{\bf y}={\bf H {\bf X} ~{\pmb \theta}}+{\bf n}.
\end{equation}
\noindent
This model is synthesis-sparse, since the image $\bf x$ is reconstructed from few objects (atoms) ${\bf X}_t$. In the proposed approach however, the synthesis dictionary $\bf X$  and the amplitudes $\pmb \theta $ are learned jointly and  iteratively through {analysis}-based priors using redundant wavelet dictionaries. Because bright sources may create artefacts  spreading all over the dirty image, objects $ {\pmb \theta}_t{\bf X}_t$ with the highest intensities are estimated at first, hopefully enabling the recovery of the faintest ones at last.
\subsection{\textbf{I}sotropic \textbf{U}ndecimated \textbf{W}avelet \textbf{T}ransform}
In the proposed method, each atom from the synthesis dictionary $\bf X$ will be estimated from its projection (analysis coefficients) in a suitable data representation. One possible choice for this representation, which  will be  illustrated below,  is  the Isotropic Undecimated Wavelet Transform (IUWT) \citep{Starck07}. IUWT dictionaries have proven to be efficient  in astronomical imaging because they allow an accurate modelling through geometrical isotropy and translation invariance. They also possess an associated fast transform.

\begin{figure}[ht]
\centering
 \includegraphics[width=1\hsize]{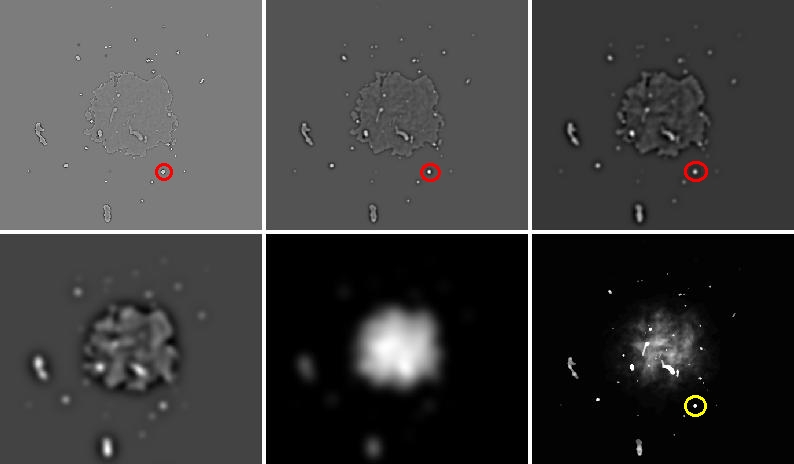}
   \caption{\small Bottom right: a galaxy cluster model image, with a galaxy circled in yellow. From top left to bottom middle: IUWT analysis coefficients ${\bf w}^\top_{(1)},\hdots,{\bf w}^\top_{(4)}$  and ${\bf a}^\top_{(4)}$ of the galaxy cluster up to the dyadic scale $J=4$. The red circles show the most significant part of the galaxy's signature in the analysis coefficients.}
  \label{dec}
   \end{figure}

We now recall some principles related to IUWT as they will be important to understand the proposed algorithm.
Analysing an image $\bf y$ of size $N$ with the  IUWT  produces analysis coefficients  that we note $ {\pmb \alpha} = {\bf A}^\top {\bf y}$. Those are composed of $J+1$ sets of wavelet coefficients, each set of the same size as the image (see Fig.\,\ref{dec}) and $J\leq \log_2 N -1$ is an integer representing the number of scales of the image decomposition. Formally  ${\pmb \alpha}$ can be written as ${\pmb \alpha} = [{\bf w}^\top_{(1)},\hdots,{\bf w}^\top_{(J)},{\bf a}^\top_{(J)}]^\top$, where the $\{{\bf w}_{(j)}\}_{j=1}^{J}$ are wavelet coefficients (for which  $j=1$ represents the highest frequencies)  and ${\bf a}_{(J)}$ is a smooth coefficients set. 
Importantly, the data $ {\bf y} $ can be recovered by $ {\bf y}=\bf S{{\pmb \alpha}} $, where  ${\bf{S}}$
is the IUWT-synthesis dictionary corresponding to ${\bf{A}}$.
   
An interesting feature of the IUWT is that astrophysical sources yield very specific signatures in its analysis coefficients. As an illustration, Fig.\,\ref{dec}  highlights a galaxy in the original image of a model galaxy cluster (see below for a more detailed description) with the  analysis coefficients generated by this galaxy (inside the red circles). The ``fingerprint'' let by this  object is clearly visible in the first three scales. This suggests that each source can in principle be associated to a set of few coefficients (w.r.t. the number of pixels $N$) which capture the source signature at its natural scales.  Conversely, we may try to reconstruct the sources from the sparse set of corresponding analysis coefficients.

This is the strategy followed below and it actually requires two steps. First, obviously, the image which we must consider to identify the sources is the dirty image,
 which is noisy. This means that all analysis coefficients  $  {\pmb\alpha} $ are not genuinely related to astrophysical information and some of them should be discarded as noise. Using standard procedures  \citep[see e.g.][]{Starck11}, the noise level can be estimated scale by scale from the wavelets coefficients using a  robust Median Absolute Deviation (MAD) \citep{Johnstone71} for instance.  The resulting \textit{significant} analysis coefficients, which we denote by $  \tilde{\pmb\alpha} $, are obtained from the analysis coefficients $  {\pmb\alpha} $  scale by scale, by leaving untouched the coefficients larger than the significance threshold and setting the others to $0$.

Second, we need a procedure that will estimate which fraction of the significant analysis coefficients $ \tilde{\pmb\alpha}$ characterizes the brightest source(s) (because we want to remove them to see what is hidden in the background). We call  this step object identification and describe it below. We will then be in position to present the global reconstruction algorithm. 
\subsection{Object identification}
The brightest object in the dirty image will be defined from a signature defined by a subset of significant coefficients $  \tilde{\pmb\alpha} $, which we denote by ${{\pmb\alpha}^{max}}$. The starting point to obtain the brightest object is to locate the most significant analysis coefficient. The IUWT analysis operation  ${\bf A}^\top \bf y={\bf A}^\top \bf H \bf x$  may be seen as the  scalar product between the atoms $\{{\bf A}_k\}_{k=1}^{N\times(J+1)}$ and ${\bf Hx}$, or equivalently of $\{{{\bf H}^\top {\bf A}_k}\}_{k=1}^{N\times(J+1)} $  with $\bf{x}$.   
 We denote  the pixel of maximal correlation score between ${\bf x}$ and one of the convolved dictionary atoms  $\{{{\bf H}^\top {\bf A}_k}\}_{k=1}^{N\times(J+1)} $  by:
\begin{equation}
\label{bright}
{k^{max}}={\displaystyle\rm arg}\max_{{k\in [1,(N+1)\times J]}}\frac{{\bf{A}}_k^\top {\bf Hx}}{\Vert {\bf{A}}_k^\top {\bf H} \Vert_2 }={\displaystyle\rm arg}\max_{{k\in [1,(N+1)\times J]}}\frac{\tilde{\pmb\alpha}_k}{\Vert {\bf{A}}_k^\top {\bf H} \Vert_2 }.
\end{equation}
This normalization ensures that if $\bf{x}$ was pure noise,  ${k^{max}}$ would pick-up all atoms with the same probability. The third part of the equation is indeed valid only if $\tilde{\pmb\alpha}$ contains nonzero coefficients. 
Let also $\alpha^{max}$ be the wavelet coefficient at the pixel position ${k^{max}}$ defined by \eqref{bright}, and $ j^{max} $ be the corresponding scale.

To formalize the object identification strategy, we need now two definitions from multiscale analysis \citep{Starck11}.
First, a set of  spatially-connected nonzero analysis coefficients at the same dyadic scale $ j$ will be called a \textit{structure} and will be denoted by ${\bf{s}}$ (this vector is thus a set of contiguous analysis coefficients). Typically, the red circles of Fig.\,\ref{dec} encircle instances of structures.
Second, an  \textit{object} will be characterized by a set of structures leaving at different scales and connected from scale to scale. Typically, the structure in the red circles of   Fig.\,\ref{dec} would be connected from scales $j=3$ to scale $j=1$ because they are ``vertically aligned'' (more precisely, the position of the maximum wavelet coefficient of the structure  at the scale $j-1$ also belongs to a structure at the scale $j$). In the case of Fig.\,\ref{dec}, the structures associated to this object correspond to only one source $-$ the circled galaxy in Fig.\,\ref{dec}.

To estimate  the whole   fingerprint of the  brightest  object (say, the circled galaxy),  we proceed as follows. First, we identify $k^{max}$ and  the structure ${\bf s}^{{max}}$, at the scale $ j^{max} $, to which belongs the pixel position ${k^{max}}$. The other structures of this object are searched  only at lower scales  ($j=1,\hdots,j^{max}-1$), where its finest details live significantly. The resulting set of connected structures  constitutes  the significant coefficients  identifying the signature of  the brightest object in the data. These coefficients are stored in a sparse vector ${{\pmb\alpha}^{max}}$ of dimension $N\times(J+1)$. 

Of course, instead of detecting and using only the most significant coefficient $k^{max}$ in $ \tilde{\pmb\alpha} $, it can be more efficient to select a fraction of the largest coefficients  at the scale $ j^{max} $ \citep[see e.g.][]{stomp}. In this case, the algorithm captures simultaneously structures  corresponding to other sources which have  intensities and natural scales that are similar to the brightest object defined only by $k^{max}$ and its associated structures. In our algorithm,  structures of the scale $j^{max}$ are allowed to have their maximum wavelet coefficient  as low as $\tau\times \alpha^{max}$, where $\tau$ is a tuning parameter, to be selected. { The joint estimation of these objects reduces significantly deconvolution artefacts since their sidelobes in the data are taken into account simultaneously. The choice of $\tau$  within a wide range (e.g. $[0.6 ~0.9])$ does not affect the final results significantly. Only small values of $\tau$ (say $0.1$) can lead to convergence issues as the fainter  objects are dominated by the brighter ones. }
In this case, ${{\pmb\alpha}}^{max}$ will capture the signature of several sources in the dirty image.  The detailed description of the resulting object identification strategy is given by \textbf{Algorithm 1}.

In this algorithm, the significant analysis coefficients ${\pmb\alpha}^{max}$ are found as in Sect. 5.1, as well as the maximally significant coefficients ${\alpha}^{max}$, its position $k^{max}$ and scale $j^{max}$ (Steps 1. and 2.). In step 3., we find and label all structures at scales smaller or equal to $j^{max}$, and we collect their number ${\pmb \mu}_j$ per scale $j$ in ${\pmb\mu}=[{\pmb \mu}_1..{\pmb \mu}_{j^{max}}]$. We store in step 5. the position of the maximum wavelet coefficient of the $i^{th}$ structure of scale $j$ in the entry ${\bf K}_{ij}$ of a matrix $\bf K$. The value of this coefficient is then ${\pmb\alpha}_{{\bf K}_{ij}}$. Step 6. involves a recursive loop. Its purpose is to look for each significant structure ${\bf{s}}^{i,j^{max}}$ at the scale $j^{max}$, whether there is a structure  at the scale $j^{max}-1$ whose maximum is vertically aligned with it. Such a structure is included in ${\pmb\alpha}^{max}$. The process is repeated for this structure at the lower scale. This process creates a tree of significant structures which describes the object's signature.

Once the signature of a bright object has been obtained, the object is deconvolved using ${{\pmb \alpha}}^{max}$ by solving approximately the following problem:
\begin{equation}
\label{critere0}
\hat{\bf z}={\displaystyle\rm arg}\min_{\bf z}\parallel{\pmb \alpha}^{max} -{\bf D}{\bf A}^\top{\bf Hz}\parallel^2_2, \text{\;\;s.t.\;\;} {\bf z} \geq\bf 0,  
\end{equation}
where $ \bf D $ maps the analysis coefficients to the nonzero values of ${{\pmb \alpha}}^{max}$ and $\bf z\geq 0$ means that all components of $\bf z$ are nonnegative.   $ \bf D $ is formally a diagonal matrix of size $(N\times(J+1), N\times(J+1)) $ defined by $  {\bf D}_{kk}=1$ if  ${{{\pmb \alpha}}^{max}_{k}}>0 $ and $0$ otherwise. 

\subsection{The MORESANE algorithm}
These ideas lead to the following iterative procedure.
At each iteration $i$, we identify a sparse vector ${\pmb \alpha}^{(i)}$, using \textbf{Algorithm 1}, that contains the signature of the brightest object in the residual image within a range controlled by $\tau$. The flux distribution of this object (that may correspond to several sources belonging to the same class in term of flux and angular scales) is estimated at each iteration $i$  as one object ${\bf z}^{(i)}$ {\color{black} using the extended Conjugate Gradient \citet{Biemond90} as described in \textbf{Algorithm 2}. In the Conjugate Gradient, since the conjugate vector and the estimate are no longer orthogonal due to the nonlinear projection on the positive orthant,  a line search method must be deployed to estimate the stepsize $\delta$.}
The estimated synthesis atom corresponding to  ~~${\bf z}^{(i)}$ is simply $\widehat{{\bf{X}}}_i=\frac{ {\bf z}^{(i)}}{\| {\bf z}^{(i)}\|_2}$ and~~ ${\widehat{\pmb{\theta}}}_i=\| {\bf z}^{(i)}\|_2$. The influence of this object can be removed from the residual image by subtracting ~~${\bf{H}}{\bf z}^{(i)}=\widehat{\pmb{\theta}}_i{\bf{H}}\;\widehat{ {\bf{X}}}_i$.
{However,} the complete removal of the bright sources contribution at each iteration could create artefacts in the residual image. Those are caused by an over-estimation of the bright contributions, which in turn can  impede the recovery of the faint objects. This  fact is reminiscent of issues regarding proper scaling of the stepsize in descent algorithms and of  CLEAN loop factor \citep{Hogbom74}. To provide a less aggressive and more progressive attenuation of the bright sources' contribution, we have introduced in MORESANE a loop gain $\gamma$ as in CLEAN. {{Values of $\gamma$ that are close to 1 lead to instability in the convergence. On the other hand, values that are too small lead to very slow convergence. We found that $\gamma \in[0.1 ~0.2]$ is a good compromise}}.  The version with  $\gamma$ is presented in \textbf{Algorithm 3}. Note that the formal number of objects may become significantly large {\color{black}when using the factor $\gamma$}.

In the reconstruction of specific examples (see the next section), it appears that when  large features are deconvolved at first, they somewhat capture the contribution of smaller sources, which are then not accurately restored during the subsequent iterations at small sales (they incur significant artefacts, in particular on the border of the sources). Therefore, we have opted for a general strategy (described in \textbf{Algorithm 4}) where \textbf{Algorithm 3} is run iteratively for {\color{black} $J=J_{PSF}$ up to  $\log_2 N-1$, where $J=J_{PSF}$ is the scale corresponding to the highest correlation of the PSF with the analysis dictionary  ${\bf A}^T$}. As the considered number of scales become larger, we also include all smaller scales in the dictionary, because small structures may become significant once the contributions of other sources have been removed.  At iteration 1, the input of \textbf{Algorithm 3} is the dirty map $\bf{y}$ and at the subsequent iterations {\color{black}($J>J_{PSF}$)} the input is the final residuals produced by \textbf{Algorithm 3 } at the previous iteration (with $J-1$ scales). Iterations may stop before $J$ reaches $\log_2 N-1$ if no significant wavelet coefficients are detected at some point. 
\newpage

{
\noindent\rule{9.1cm}{1pt}\\
{\small
\textbf{Algorithm 1} Object Identification\\
\noindent\rule{9.3cm}{1pt}\\
\noindent
\textbf{Input}: ${\pmb\alpha}$, $\tau$.\\
\textbf{Output}: ${\pmb\alpha}^{max}$.\\
$\bullet$ Identify the significant analysis coefficients $\tilde{\pmb\alpha}$ as in Sect. 5.1.\\
$\bullet$ I
dentify ${k^{max}}$ \eqref{bright} and its corresponding  $\alpha^{max}$  and $ j^{max} $.\\
$\bullet$ Find and label all structures of $\tilde{\pmb\alpha}$ at scales $j=1$ to $j^{max}$.\\
$\bullet$ Determine ${\pmb\mu}$ (${\pmb\mu}_j$ is the number of structures at a scale $j$).\\
$\bullet$ Determine the pixel position of the maximum wavelet coefficient of the $i^{th}$ structure (denoted by  ${\bf{s}}^{i,j}$) at each scale $j$ ($j=1$ to $j^{max}$), and store it in a matrix entry ${\bf K}_{ij}$, the value of its wavelet coefficient is  $\tilde{\pmb\alpha}_{{\bf K}_{ij}}$.\\
$\bullet$ \textbf{for} $i=1$ \textbf{to} ~ ${\pmb \mu}_{j^{max}}$
\begin{itemize}
\item[{1.}] \textbf{if} $\tilde{\pmb\alpha}_{{\bf K}_{ij^{max}}}\geq \tau\times\alpha^{max}$

1.1. Add ${\bf s}^{i,j^{max}}$ to ${\pmb\alpha}^{max}$.\\
1.2. Initialize $\ell=1$.\\
1.3. \textbf{for} $t=1$ \textbf{to} ~ ${\pmb\mu}_{j^{max}-l}$ 

\textbf{~~~~~~~~~~~~if} ${\bf K}_{t{j^{max}-l}}$ is in the support of ${\bf s}^{i,j^{max}-l+1}$

{~~~~~~~~~~~~~~~1.3.1}. Add  ${\bf s}^{t,{j^{max}-l}}$ to ${\pmb\alpha}^{max}$.

~~~~~~~~~~~~~~~{1.3.2}. Set $\ell=\ell+1$.

~~~~~~~~~~~~~~~{1.3.3}. Repeat {1.3.} until $\ell=j^{max}$. 

\textbf{~~~~~~~~~~~~end if}.

1.4. \textbf{end for}.
\item[{2.}] \textbf{end if}.
\end{itemize}
$\bullet$ \textbf{end for}.\\
\noindent
\noindent\rule{9.1cm}{1pt}\\
\textbf{Algorithm 2} Conjugate Gradient method: Minor cycle\\
\noindent\rule{9.1cm}{1pt}\\
\noindent
\textbf{Input}: ${\pmb\alpha}$, $ {\bf D}$, $L_{itr}$, $\bf H$,  $J$, {\color{black}$\epsilon$}.\\
\noindent
\textbf{Output}: deconvolved objects $\hat{\bf z}$.\\
\noindent
$\bullet$ Initialize  $\ell=0$, iteration index, ${\bf r}^{(0)}= {\bf S}{\pmb \alpha}$, residual image, ${\bf v}^{(0)}={\bf r}^{(0)}$, gradient, $\hat{\bf z}^{(0)}=\bf 0$, ${\bf D}{\bf A}^\top{\bf H}\equiv{\bf W}$. \\
\noindent
$\bullet$ \textbf{while}  $\ell< L_{itr}$ \textbf{do}
\begin{enumerate}
\item ${\bf z}^{(\ell+1)}={\cal{P}_+}({\bf z}^{(\ell)}+\delta {\bf v}^{(\ell)})$, ${\cal{P}_+}$ is a projection operator on $\mathbb{R}^N_+$, {\color{black} and the stepsize  $\delta$ is calculated using a line search method.}
\vspace{2mm}
\item ${\bf r}^{(\ell+1)}={\bf r}^{(\ell)}-\delta ~{\bf SWv}^{(\ell)}$.
\vspace{2mm}
\noindent
\item {\color{black}$\beta=\frac{\langle {\bf r}^{(\ell+1)}-{\bf r}^{(\ell)},{\bf r}^{(\ell+1)}\rangle}{\langle{\bf r}^{(\ell)},{\bf r}^{(\ell)}\rangle}$.}
\vspace{2mm}
\noindent
\item ${\bf v}^{(\ell+1)}={\bf r}^{(\ell+1)}+\beta{\bf v}^{{(\ell)}}$.
\item Set $\ell=\ell+1$.
\end{enumerate}
Iterations stop if
{\color{black}$\dfrac{\parallel{\bf z}^{(\ell+1)}-{\bf z}^{(\ell)}\|_2}{\parallel{\bf z}^{(\ell)}\|_2}<\epsilon$.\\}
$\bullet$ \textbf{end while.}\\
\noindent
\noindent\rule{9.1cm}{1pt}\\
\textbf{Algorithm 3} Objects estimation\\
{
\noindent\rule{9.1cm}{1pt}\\
\noindent
\textbf{Input}: ${\bf r}$, $\bf H$, $\tau$, $\gamma$, $J$, {\color{black}$\varepsilon$}, $N_{itr}$.\\
\noindent
\textbf{Output}: $\widehat{\bf  X}$, $\widehat{\pmb{\theta}}$, residual $\bf r$ .\\
\noindent
$\bullet$ Initialize  $i=0$, major iteration index, ${\bf r}^{(0)}=\bf r$, $\bf \widehat X=0$, $\widehat{\pmb{\theta}}=\bf 0$.\\
\noindent
$\bullet$ Determine ${{\pmb \alpha}}^{(0)}$, the sparse analysis vector corresponding to  the  brightest objects in ${\bf r}^{(0)}$ using \textbf{Algorithm 1}, and thus $ {\bf D}^{(0)} $. \\
\noindent
$\bullet$ \textbf{while}   ${\pmb \alpha}^{(i)} \neq{\bf{0}}$ and $i<N_{itr}$ \textbf{do} 

\begin{enumerate}
\item \textbf{Analysis} based deconvolution step: Compute \\
\begin{equation*}
\hat{\bf z}^{(i)}={\displaystyle\rm arg}\min_{\bf z}\parallel{\pmb \alpha}^{(i)} -{\bf D}^{(i)}{\bf A}^\top{\bf Hz}\parallel^2, \text{\;\;s.t.\;\;} {\bf z} \geq\bf 0, \text{using \textbf{Algorithm 2}}
\end{equation*}
\vspace{1mm}
\item $\widehat{{\bf{X}}}_i=\frac{ \hat{\bf z}^{(i)}}{\| \hat{\bf z}^{(i)}\|_2}$ and ${\widehat{\pmb{\theta}}}_i=\| {\hat{\bf z}}^{(i)}\|_2$
\vspace{1mm}
~~\item Update $ {\bf r}^{(i+1)}={\bf r}^{(i)}-\gamma{\widehat{\pmb{\theta}}}_i{\bf H}\widehat{\bf X}_i.$
\item Determine  ${{\pmb \alpha}}^{(i+1)}$, using \textbf{Algorithm 1}, and $ {\bf D}^{(i+1)} $.
\item Set $i=i+1$. 
\end{enumerate}
\noindent
Iterations stop if
{\color{black}$\dfrac{\parallel \sigma_{{\bf r}^{(i)}}-\sigma_{{\bf r}^{(i-1)}}\|_2}{\parallel\sigma_{{\bf r}^{(i-1)}}\|_2}<\varepsilon$, where $\sigma_{{\bf r}^{(i)}}$ is the standard deviation of the residual ${\bf r}^{(i)}$. \\
}
$\bullet$ \textbf{end while.}\\
\noindent\rule{9.1cm}{1pt}\\
}
\newpage
\noindent\rule{9.1cm}{1pt}\\
\textbf{Algorithm 4} MORESANE\\
\noindent\rule{9.1cm}{1pt}\\
\noindent
\textbf{Input}: $\bf y$, $\bf H$, $\tau$, $\gamma$, $N_{itr}$.\\
\noindent
\textbf{Output}: reconstructed image $\bf \hat x$ .\\
\noindent
$\bullet$ {\color{black} Compute $J_{PSF}$ corresponding to the scale of the highest correlation of the PSF with the IUWT-analysis dictionary}.\\
$\bullet$ Initialize  {\color{black}$J=J_{PSF}$}, number of scales for the IUWT-decomposition, ${\bf r}^{(0)}= \bf y$.\\
$\bullet$ \textbf{while} $J< \log_2 N-1$
\begin{enumerate}
\item Determine $\widehat{\pmb\theta}_{(J)}$ and $\widehat{{\bf X}}_{(J)}$ using \textbf{Algorithm 3}.
\item Update dictionary $\widehat{\bf X}=[\widehat{\bf X}~~\widehat{\bf X}_{(J)}]$.
\item Update weights  $\widehat{\pmb{\theta}}=[\widehat{\pmb{\theta}}~~\widehat{\pmb{\theta}}_{(J)}]$.
\item Update residual ${\bf r}^{(J)}= {\bf r}^{(J-1)}-\gamma \widehat{\bf X}_{(J)}~\widehat{\pmb{\theta}}_{(J)}$.
\item Set $J=J+1$.
\end{enumerate}
iterations stop if $\widehat{\bf X}_{(J)}=\bf 0$.\\
$\bullet$ \textbf{end while.}\\
$\bullet$ \textbf{Synthesis} step: $\hat{\bf x}=\gamma\widehat{\bf X}~{\widehat{\pmb{\theta}}}.$\\
}
\noindent\rule{9.1cm}{1pt}

\section{Application of MORESANE and the benchmark algorithms} 
\label{simulations}
In this section, we evaluate the performance of the deconvolution algorithm MORESANE in comparison with the existing benchmark algorithms. We provide two families of tests. In the first scheme, we apply MORESANE on realistic simulations of radio interferometric observations. The results 
 are compared to those obtained by the classical CLEAN-based approaches (H\"ogbom CLEAN and Multi-scale CLEAN) and the deconvolution compressive sampling method developed in \citeyear{Li11} by \citeauthor{Li11} (IUWT-based CS method in the following). In the second scheme, we apply MORESANE on simplified simulations of radio data, where the considered uv-coverage is a sampling function with 0 and 1 entries in order to compare MORESANE with the SARA algorithm developed in \citeyear{Carrillo12} by \citeauthor{Carrillo12}. The published code of the latter is currently applicable only to a binary uv-coverage and could not therefore be applied to the first set of our simulations. 
 
The simulated data presented in this paper concern two kinds of astrophysical sources containing both complex extended structures and compact radio sources.  We firstly consider a model of a galaxy cluster. Similarly to observed galaxy clusters \citep[see e.g. Fig. 1 in][]{2006A&A...460..425G}, the adopted model hosts a wide variety of radio sources, such as: a) point-like objects, corresponding to unresolved radio galaxies; b) bright and elongated features related to tailed radio-galaxies, which are shaped by the interaction between the radio plasma ejected by an active galaxy and the intra-cluster gas observed in X-rays \citep[e.g.][]{2002ASSL..272..163F}; and c) a diffuse radio source, so called ``radio halo'', revealing the presence of relativistic electrons (Lorentz factor $\gamma$ >> 1000) and weak magnetic fields ($\sim \mu$Gauss) in the intra-cluster volume over Mpc scales \citep[e.g.][]{2008SSRv..134...93F}. So far, only a few tens of clusters are known to host diffuse radio sources \citep[see e.g.][for recent reviews]{2012A&ARv..20...54F,2014IJMPD..2330007B}, which are extremely elusive due to their very low surface brightness. The model cluster image adopted in this paper (courtesy M. Murgia and F. Govoni) has been produced using the FARADAY tool \citep{2004A&A...424..429M} as described in Ferrari et al. (2014, in prep.). We secondly analyze the {\color{black}toy} image of an HII region in M31 that has been widely adopted in most of previous deconvolution studies \cite[e.g.][ see also \url{http://casaguides.nrao.edu/index.php?title=Sim_Inputs}]{Li11, Carrillo12} due to its challenging features, i.e. high signal-to-noise and spatial dynamic ranges. 
 In the following, all the maps are shown in units of Jy/pixel.
\subsection{Results on simulations of realistic observations} 

We simulate observations performed with MeerKAT. The radio telescope, currently under construction in South Africa, will be one of the main precursors to the SKA. By mid 2017, MeerKAT will be completed and it will then be integrated into the mid-frequency component of SKA Phase 1 (SKA1-MID). In its first phase MeerKAT will be optimized to cover the L-band (from $\approx$ $1$ to $1.7$ GHz). It will be an array of 64 receptors, among which $48$ will be concentrated in a core area of approximately $1$ km in diameter, with a minimum baseline of $29$ m (corresponding to a detectable largest angular scale of about $25$ arcmin at $1.4$ GHz). The remaining antennas will be distributed over a more extended area, resulting in a distance dispersion of $2.5$ km and a longest baseline of 8 km (corresponding to maximum achievable resolution of about 5.5 arcsec at 1.4 GHz). Both the inner and outer components of the array will follow a two-dimensional Gaussian uv-distribution, that produces a PSF whose central lobe can be nicely reproduced with a Gaussian shape.

Our test images are shown in the top panels of Fig.\,\ref{data}. Their brightness ranges from 0 to 4.792 $\times 10^{-5}$ Jy/pixel and from -2.215 $\times 10^{-9}$ to 1.006 Jy/pixel for the cluster and M31 cases, respectively (with 1 pixel corresponding to 1 arcsec). The center of the maps is taken to be located at RA=0 and Dec=-40 degrees (note that MeerKAT will be located at latitude $\sim$-30 degrees).  To simulate realistic observations, we use the MeqTrees package \citep{Smirnov10}. We consider a frequency range from 1.015 GHz to 1.515 GHz, with an integration time of 60 seconds and a total observation time of 8 hours. We adopt a robust weighting scheme (with a Briggs robustness parameter set to 0) and a cell size of 1 arcsec, corresponding to $\sim 1/5$ of the best angular resolution achievable by MeerKAT. The resulting standard deviation of the noise in the simulated maps is 1.73 $\times 10^{-6}$ Jy/pixel. The simulated image sizes are selected to be $2048 \times 2048$ pixels, corresponding to roughly 1/3 of the primary beam size of MeerKAT ($\approx$ 1.5 deg at 1.4 GHz). The sky images shown in Fig.\,\ref{data}, originally both of size $512\times 512$ pixels, are padded with zeros in their external regions. 

{\color{black} CLEAN based approaches are performed directly on the continuous visibilities using  the ``lwimager'' software implemented in MeqTrees, a stand-alone imager based on the CASA libraries and providing CASA-equivalent implementations of various CLEAN algorithms. Whereas, both MORESANE and IUWT-based CS, written in MATLAB, take for entries the PSF and the dirty image and  work on the gridded visibilities using the Fast Fourier Transform. }
\begin{figure}[ht]
   {\centering
  
  \includegraphics[width=0.5\hsize]{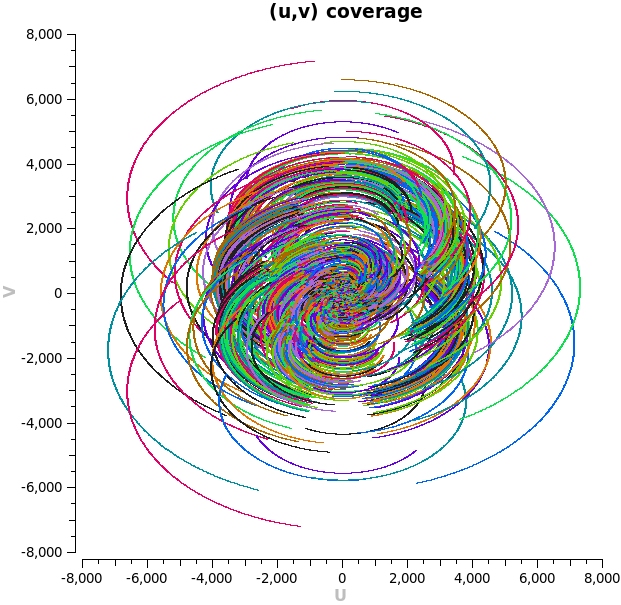}
\includegraphics[width=0.45\hsize]{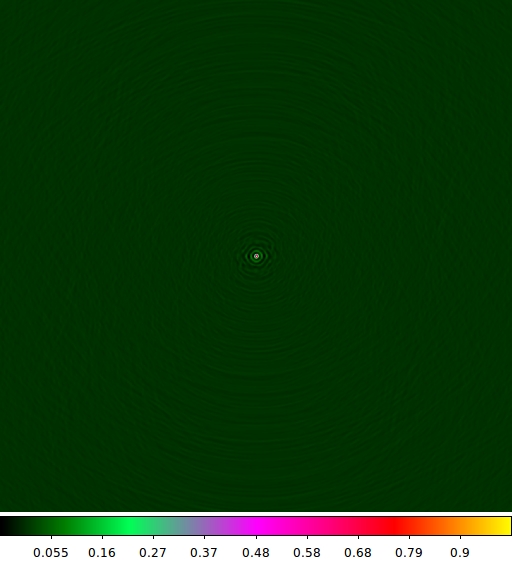}
     }           
           \caption{\small Left: $(u,v)$ coverage of MeerKAT for 8 hours of observations, colors correspond to the same baseline, right: its corresponding PSF.}  
           \label{uvcoverage}
   \end{figure}
   
\begin{figure*}[ht]
\centering
\includegraphics[width=0.24\hsize]{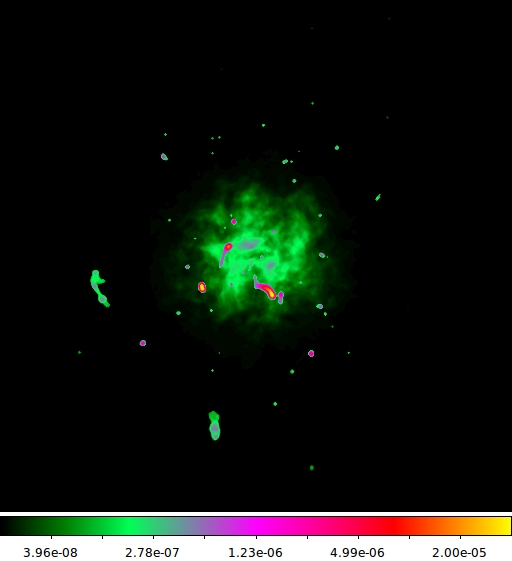}
\includegraphics[width=0.24\hsize]{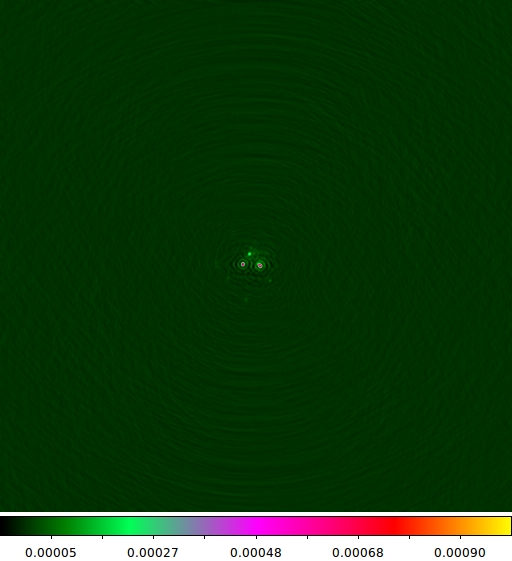}
\includegraphics[width=0.24\hsize]{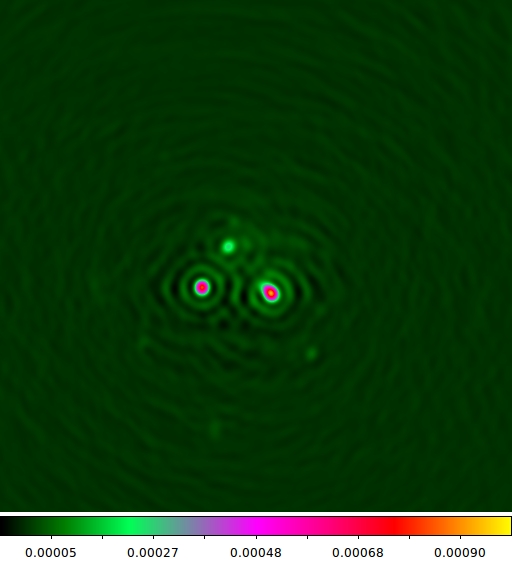}

\includegraphics[width=0.24\hsize]{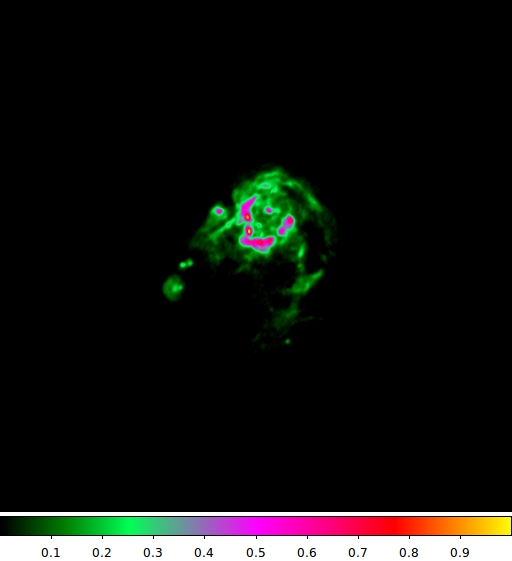}
\includegraphics[width=0.24\hsize]{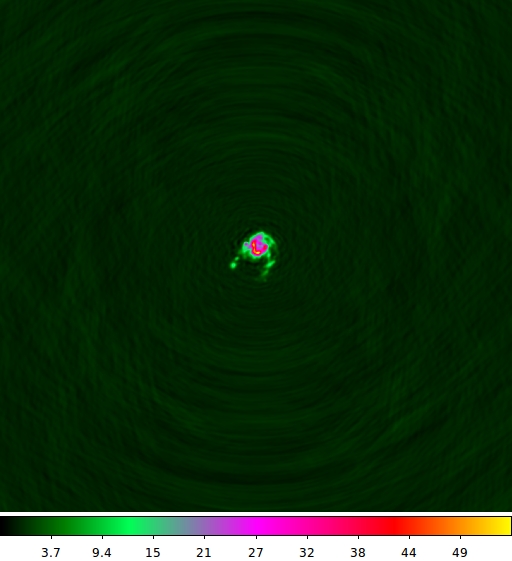}
\includegraphics[width=0.24\hsize]{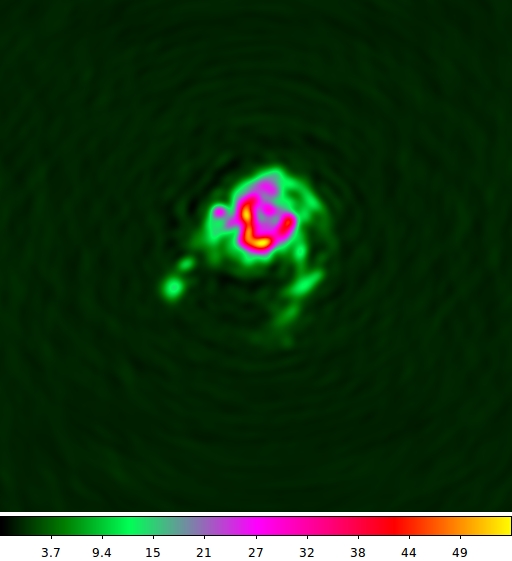}

 \caption{\small {Top:  simulated galaxy cluster data. Bottom: simulated M31 data. From left to right:  input images of size $512 \times 512$ pixels (shown respectively in log scale for the galaxy cluster and linear scale for M31),  dirty images of size $2048 \times 2048$ pixels and zoom on their  central regions of size $512 \times 512$ pixels.}}
     \label{data}    
\end{figure*}

The uv-coverage and corresponding PSF of the simulated observations are shown in Fig.\,\ref{uvcoverage}. The main lobe of the PSF is approximated by a Gaussian clean beam (10.5 arcsec $\times$ 9.9 arcsec, P.A.=-28 deg). The resulting dirty images provided by MeqTrees are shown in the middle  panel of Fig.\,\ref{data}. 
   
Due to the important dynamic range ($\approx$1:10000) of the cluster model map, the diffuse radio emission of the radio halo in the dirty map is completely buried into the PSF side lobes of bright sources (see {\color{black}top panel} of Fig.\,\ref{data}). To perform the deconvolution step with MORESANE, we consider the following entries.  The gain factor $\gamma$ that controls the decrease of the residual is set to {$\gamma=0.2$}. The parameter $\tau$ that controls the number of detected objects per iteration is set to $\tau=0.7$ and the maximum number of iterations to $N_{itr}=200$ and {\color{black}$\varepsilon=0.0001$}. {For the wavelets denoising, we use $4\sigma$ clipping.} For the minor cycle,  we fix the maximum number of iterations {\color{black}in the extended Gradient Conjugate $L_{itr}$ to $50$, (note that tests have shown that convergence is usually reached before) and the precision parameter $\epsilon$ to 0.001}.  MORESANE stops at $J=7$. For both H\"ogbom and Multi-scale CLEAN tests,  we set $\gamma = 0.2$, {the threshold to $3\sigma$} and the maximum number of iteration to $N_{itr}=10000$. More specifically to the Multi-Scale CLEAN, we use $7$ scales $[0,2,4,8,16,32,64]$ and {$\gamma=0.2$}. For the IUWT-based CS, we use its re-weighted version implemented in MATLAB (found at \url{https://code.google.com/p/csra/downloads}). We set the level of the IUWT-decomposition to { $6$}, the threshold to $5$ percent of the maximum value in the Fourier transform of the PSF, { the regularization parameter $\lambda=10^{-8}$, the threshold to $3\sigma$ and the maximum number of iteration $N_{itr}=50$}.

The dirty image corresponding to M31 is displayed in {\color{black}the bottom panel} of Fig.\,\ref{data}. The source is completely resolved and above the noise level. To deconvolve it, we set the parameters of MORESANE to $\tau=0.7$, $\gamma=0.2$,  $N_{itr}=200 $, {\color{black}$\varepsilon=0.0005$} and  $5\sigma$ clipping on the wavelet domain. {\color {black}In the minor loop, we set the precision parameter $\epsilon =0.01$.} MORESANE stops at $J=7$. In the case of the H\"ogbom CLEAN we set: $\gamma=0.2$, $N_{itr}=10000$ and a {3$\sigma$ }threshold. For Multi-Scale CLEAN we adopt:  $\gamma=0.2$, $N_{itr}=10000$, a $3\sigma$ threshold and $7$ scales ($[0,2,4,8,16,32,64]$). Finally, the parameters set for the IUWT-based CS method are: $7$ scales, the threshold to $5$ percent of the maximum value in the Fourier transform of the PSF, $\lambda=10^{-4}$,  the threshold to {$3\sigma$} and the maximum number of iteration { $N_{itr}=50$.}

To quantify numerically the quality of the image recovery by MORESANE with respect to the benchmark algorithms, in terms of fidelity and dynamic range,  we use two indicators described in the following.
\begin{enumerate}
\item [i)] The Signal to Noise Ratio ($SNR$) is defined as the ratio of the standard deviation $\sigma_{\bf x}$ of the original sky to the standard deviation $\sigma_{\bf \hat x - x}$ of the estimated model from the original sky.
\begin{equation}
\label{SNR}
SNR=20 \log_{10} \frac{\Vert{\bf x}\Vert_2}{\Vert {\bf \hat x -x}\Vert_2}.
\end{equation}
\noindent 
\end{enumerate}
The CLEAN algorithm provides a very poor representation of the original scene, since it is assumed to be only composed of point sources. Therefore, the $SNR$ of the CLEAN model is very low inherently. For a more reliable evaluation of the image recovery given by the four algorithms, we  use the $SNR$ metric on the model images convolved with a \textit{clean beam}. The latter is usually a two dimensional elliptical Gaussian that fits the primary lobe of the PSF.  These images are considered to be more reasonable from the astrophysics point of view, especially in the case of the CLEAN algorithm and its variants. Hereafter, we call a \textit{beamed} image, an image convolved with a clean beam. 

\begin{figure*}[ht]
\centering
\includegraphics[width=0.275\hsize]{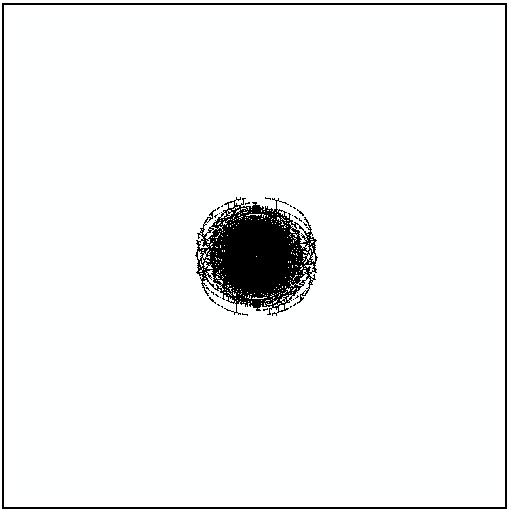}
\includegraphics[width=0.25\hsize]{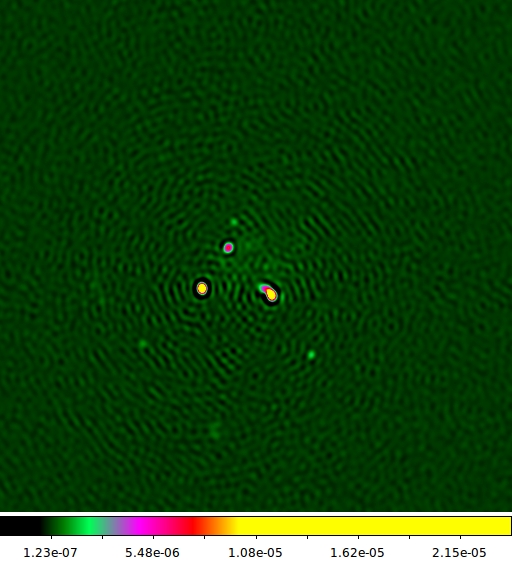}
\includegraphics[width=0.25\hsize]{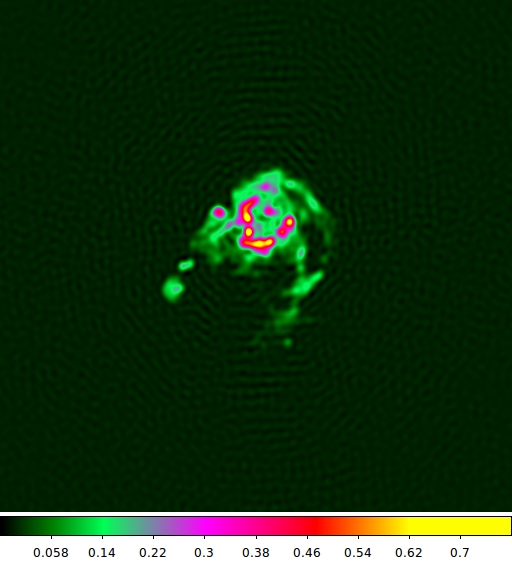}
\caption{{\color{black}Left considered uv-coverage, middle the dirty image corresponding to the galaxy cluster and right the dirty image corresponding to the M31.}}
\label{simforsara}
\end{figure*}
Radio astronomers usually refer to the \textit{restored map} ${\bf \tilde y}$ given by:
\begin{equation}
\label{rest}
{\bf \tilde y}=\bf B\hat x+r.
\end{equation}
where $\bf B$ is the convolution matrix by the clean beam and $\bf r$ is the residual image of the deconvolution. 
\begin{enumerate}
\item [ii)]The Dynamic Range metric ($ DR $) is defined in \cite*{Li11}, as the ratio of the peak brightness of the restored image to the standard deviation $\sigma_{\bf r}$ of the residual image,
\begin{equation}
DR=\frac{||\tilde{\bf{y}}||_\infty}{\sigma_{\bf r}}.
\end{equation}
\end{enumerate}

Fig.\,\ref{halo_rslts} shows the deconvolution results obtained on the galaxy cluster. The model images, the beamed images, the beamed error images  and the deconvolution residual images are displayed. From a qualitative inspection of Fig.\,\ref{halo_rslts}, MORESANE and the IUWT-based CS method provide  better approximations of the original scene than CLEAN, the morphologies of the different objects being estimated in a more accurate way. MORESANE is additionally more robust to false detections: while the two versions of CLEAN and the IUWT-based CS method detect an important number of fake components, almost all objects in the MORESANE model correspond to genuine sources when checked against the true image. 

For a more quantitative comparison between the different methods, we compare the photometry of the reconstructed models versus the true sky. In the case of the galaxy cluster, the total flux density of the true sky over the central $512 \times 512$ pixel area is $ 4.10\times 10^{-3}$ Jy. The total flux values that we get in the cases of H\"ogbom CLEAN, Multi-scale CLEAN, IUWT-based CS and MORESANE are respectively {$ 3.4\times 10^{-3},~ 3.6\times 10^{-3},~ 8.3\times 10^{-3},~{\color{black}4\times 10^{-3}}$}. We also compare the photometry pixel by pixel as shown in Fig.\,\ref{HALOR}, where we plot the estimated model images on the $y$-axis against the true sky image on the $x$-axis. In both tests, MORESANE is the method that gives better results in terms of total flux and surface brightness. MORESANE is also giving better results in terms of  $SNR$ on the beamed models introduced before (see top part of Table\,\ref{table:1}).

The results of M31 reconstruction {\color{black}confirm the better performance of MORESANE}. In Fig.\,\ref{M31}, we do not show the beamed models where the differences, unlike on the non-beamed versions, are negligible. Instead, we show both the error images $\bf x-\hat x$ (Fig.\,\ref{M31:edge-b}) and its beamed version $\bf Bx-B\hat x$ (Fig.\,\ref{M31:edge-c}). While the IUWT-based CS gives {\color{black}{a very good}} estimation of the model source, as confirmed by inspection of Fig.\,\ref{M31:edge-a} and Fig.\,\ref{M31R}, it is still less competitive than MORESANE when comparing fidelity tests and dynamic range results (bottom part of Table\,\ref{table:1}). This is strongly related to false detections. 
 The total flux of the sky image is $1495.33$ Jy. The reconstructed total flux by  H\"ogbom CLEAN, Multi-scale CLEAN, IUWT-based CS and MORESANE are respectively  {$1495,~1495.7,~1533$ and {\color{black}$1495.8$}}. Both MORESANE and CLEAN conserve very well the flux, while the high false detection rate of the IUWT-based CS method explains its higher total flux value.

\subsection{Results on simplified simulations of observations} 

In order to compare the performance of MORESANE with the algorithm SARA, we use toy simulations of radio interferometric images, where the considered uv-coverage is a binary sampling function. The latter is derived from the previously generated PSF of MeerKAT. Considering the central part of the PSF of size $512 \times 512$ pixels, Fourier samples with very low magnitude ($< 0.01$ of the maximum) are put to zero, as well as the central frequency. The remaining values are set to 1, keeping only $4\%$ of the measurements. The resulting new PSF is simply the inverse Fourier Transform of the new uv-coverage. Within this configuration, simulated radio images  corresponding to the galaxy cluster and M31 are shown in Fig.\,\ref{simforsara}. An additive white noise of standard deviation 6 $\times 10^{-8}$ Jy/pixel  is  added to the visibilities in order to mimic  a similar noise level to the previous simulations. 

The SARA algorithm has shown its superiority to the IUWT-CS based algorithm in \citet{Carrillo12}. Therefore, in this paragraph the performance of MORESANE is studied with respect to SARA only. To do so, we use the MATLAB code of the SARA algorithm (found at \url{https://github.com/basp-group/sopt}). {\color{black} Note that in this set of simulations visibilities are lying on a perfect grid. }  MORESANE results are obtained using the same parameters  as for the precedent test. 

Deconvolution results for the galaxy cluster are shown in Fig.\,\ref{haloS}. Clearly MORESANE provides a better model than SARA, as confirmed numerically in Table.\,\ref{table:2}.  The total flux density of the true sky is $ 4.10\times 10^{-3}$ Jy. The total flux values that we get in the cases of SARA and MORESANE are respectively $4.32\times 10^{-3}$ and $4\times 10^{-3}$. In the case of M31 reconstruction, SARA has proved to perform better deconvolution than MORESANE as shown in Fig.\,\ref{SM31}. The total flux of the sky image is $1495.33$ Jy and its estimated values by SARA and MORESANE are respectively  {$1496.1$ and $1496.7$}. Furthermore, SARA provides better $DR$ and $SNR$.

In SARA, very faint false components are  reconstructed all over the field. Our understanding of this effect is that the method minimizes the difference between the observed and modeled visibilities within an uncertainty range, which is defined inside the algorithm with respect to the noise level.  Small errors in the modeled visibilities give rise to weak fluctuations within the whole reconstructed image (in this case a factor of $\approx$0.01 lower than the minimum surface brightness of the source). The SARA estimated model is the one that best describes the visibilities, within an error margin and subject to an analysis-sparse regularization. This strategy results here in a residual image with very low standard deviation, despite the artefacts visible all over the field (see Fig.\,\ref{SM31log}, middle panel). 
 
The right panel of Fig.\,\ref{SM31log} shows that MORESANE model includes a weak (a factor of $\approx$0.1 lower than the minimum surface brightness of the source) fake emission at the edge locations of M31. Because  MORESANE uses  dictionaries based  on isotropic wavelets,  edges are less well preserved in the current case where the source is fully resolved, extended and significantly above the noise level. On the other hand, the MORESANE method does not produce  false detections in the field surrounding the source, because source detection (Algorithm 1) and reconstruction (Algorithm 2) are  done locally, respectively in the wavelet and image domains.

\begin{table*}[ht]
  \centering{     
\begin{tabular}{c c c c c}
\hline
\noalign{\vskip 0.1cm}    
 & H\"ogbom CLEAN & Multi-Scale CLEAN & IUWT-based CS & MORESANE \\
\noalign{\vskip 0.1cm}    
\hline 
\hline                       
\noalign{\vskip 0.1cm}    
{\bf Galaxy cluster} & & & & \\
\noalign{\vskip 0.1cm}    
\hline
\noalign{\vskip 0.1cm}    
SNR on the beamed models [dB] & 24.9012 & 27.5243 & 20.2848 &  33.44 \\      
\noalign{\vskip 0.1cm}    
\hline 
\noalign{\vskip 0.1cm}    
DR & 456.21 & 498.43 & 502.21 & 543.40\\
\noalign{\vskip 0.1cm}    
\hline    
\hline                        
\noalign{\vskip 0.1cm}    
{\bf M31} & & & & \\
\noalign{\vskip 0.1cm}    
\hline
\noalign{\vskip 0.1cm}    
SNR on the beamed models [dB] & 55.7455 & 51.6475  & 43.9001 & 59.3224 \\      
\noalign{\vskip 0.1cm}    
\hline 
\noalign{\vskip 0.1cm}    
DR [$\times 10^4$] & 1.5405 & 0.6881 & 0.4356 &  1.8541\\
\noalign{\vskip 0.1cm}    
\hline 
\noalign{\vskip 0.1cm} 
\end{tabular}
}
\caption{{\color {black}Numerical comparison between the different deconvolution results on realistic simulations}}
\label{table:1} 
\end{table*}

\begin{table}[!h]
  \centering{     
\begin{tabular}{c c c}
\hline
\noalign{\vskip 0.1cm}    
 & SARA & MORESANE \\
\noalign{\vskip 0.1cm}    
\hline 
\hline                       
\noalign{\vskip 0.1cm}    
{\bf Galaxy cluster} & &  \\
\noalign{\vskip 0.1cm}    
\hline
\noalign{\vskip 0.1cm}    
SNR on the models [dB] &13.31 &  16.34  \\      
\noalign{\vskip 0.1cm}    
\hline 
\noalign{\vskip 0.1cm}    
SNR on the beamed models [dB] & 26.17 &  29.47 \\
\noalign{\vskip 0.1cm}    
\hline 
\noalign{\vskip 0.1cm}    
DR &395.41  &397.07\\
\noalign{\vskip 0.1cm}    
\hline   
\hline                        
\noalign{\vskip 0.1cm}    
{\bf M31} & & \\
\noalign{\vskip 0.1cm}    
\hline
\noalign{\vskip 0.1cm}    
SNR on the models[dB] &23.19   &17.22  \\      
\noalign{\vskip 0.1cm}    
\hline 
\noalign{\vskip 0.1cm}    
SNR on the beamed models [dB] &47.67  & 38.81 \\
\noalign{\vskip 0.1cm}    
\hline 
\noalign{\vskip 0.1cm}    
DR [$\times 10^7$] & 1.38  &  0.0007\\
\noalign{\vskip 0.1cm}    
\hline
\noalign{\vskip 0.1cm} 
\end{tabular}
}
\caption{{\color{black}Numerical comparison SARA versus MORESANE on toy simulations}}
\label{table:2} 
\end{table}

\section{Summary and conclusions} \label{Conclusions}

In this paper, we present a new radio deconvolution algorithm -- named MORESANE (MOdel REconstruction by Synthesis-ANalysis Estimators) -- that combines complementary types of sparse recovery methods in order to reconstruct the most appropriate sky model from observed radio visibilities. A synthesis approach is used for the reconstruction of images, in which the unknown synthesis objects are  learned using analysis priors. 

The algorithm has been conceived and optimized for the restoration of faint diffuse astronomical sources buried in the PSF side lobes of bright radio sources in the field. A typical example of important astrophysical interest is the case of galaxy clusters, that are known to host bright radio objects (extended or unresolved radio galaxies) and low-surface brightness Mpc-scale radio sources \citep[$\approx \mu{\rm Jy}/{\rm arcsec}^2$ at 1.4 GHz,][]{2008SSRv..134...93F}. 

In order to test MORESANE capabilities, in this paper we simulate realistic radio interferometric observations of known images, in such a way to be able to directly compare the reconstructed image to the original model sky. Observations performed with the MeerKAT array (i.e., one of the main SKA-pathfinders, that is being built in South Africa) are simulated using the MeqTrees software \citep{Smirnov10}. We consider two sky models, including the image of an HII region in M31, which has been widely adopted in most of previous deconvolution studies, and a model image of a typical galaxy cluster at radio wavelengths, which has been produced using the FARADAY tool \citep{2004A&A...424..429M}. We then compare MORESANE deconvolution results to those obtained { by available tools that can be directly applied to radio measurement sets, i.e. } the classical CLEAN and its multiscale variant \citep{Cornwell08} and one of the novel Compressed Sensing approaches, the IUWT-based CS method by \citet{Li11}.

Our results indicate that MORESANE is able to efficiently reconstruct images of a wide variety of sources (compact point-like objects, extended tailed radio galaxies, low-surface brightness emission) from radio interferometric data. 
In agreement with the conclusions based on other CS-based algorithms \citep[e.g.][]{Li11,2014arXiv1406.7242G}, the MORESANE output model has a higher resolution compared to CLEAN-based methods (compare, e.g., the second and fourth images in the first column of Fig.\,\ref{halo_rslts}) and represents an excellent approximation of the scene injected in the simulations.

{\color{black}Results} obtained in the galaxy cluster case ( Fig.\,\ref{halo_rslts} {\color{black} and Fig.\,\ref{haloS}}), as well as the fidelity tests summarized in the top part of Table\,\ref{table:1} {\color{black} and Table\,\ref{table:2}}, clearly indicate that MORESANE provides a better approximation of the original scene than the other deconvolution methods. {\color{black}In both sets of simulations,} the new algorithm proved to be  more robust to false detections: while multi-scale CLEAN, the IUWT-based CS  {\color{black} and SARA} methods detect an important number of fake components, almost all objects in the MORESANE model correspond to genuine sources when checked against the true image. In addition, MORESANE gives  better results when comparing the correspondence between the true sky pixels and those reconstructed (see Fig.\,\ref{HALOR}). {\color{black}This proves that MORESANE is robust in the case of a noise level significantly higher than the weakest source brightness in the field}. These are a valuable results for getting an output catalogue of sources from radio maps. New radio surveys coming from SKA and its pathfinders will allow to get all-sky images at (sub-)mJy level, thus requiring extremely efficient and reliable source extraction methods \citep[][and references therein]{SPARCS13}. In addition, due to the huge data rate of the new generation of radio telescopes (300 Gigabytes per second in the case of LOFAR, that will increase by a factor of at least one hundred with SKA), observations will not be systematically stored, but data reduction will have to be completely automatized and done ``on the fly''. We plan to further develop MORESANE to automatically extract an output catalogue of sources (position, size, flux, \dots) from its reconstructed model. This would allow to easily insert our new image reconstruction method in pipelines for automatic data reduction, based also on the fact that our tests indicate that, contrarily to the IUWT-CS method, the settings of parameters of MORESANE do not need a fine tuning of the user, but can be easily optimized for generalized cases. 

The results of M31 reconstruction are less conclusive about the best deconvolution method.  {\color{black}On the realistic simulations, while the IUWT-based CS gives a very good estimation of the model source, it is still less competitive than MORESANE when comparing fidelity tests and dynamic range results due to the high rate of false model components.However, on the toy simulations of M31,  SARA outperformed MORESANE with higher dynamic range and fidelity. In the considered  M31 toy model, the source is fully resolved  and with a noise level lower than  the intensity of its weakest component. We stress here that in true observations,  these criteria are met only when observing bright sources with long exposure times and within small filed of views. 
}

These results are extremely encouraging for the application of MORESANE to radio interferometric data. Further developments are planned, including {comparing our tool to other existing algorithms that are, for the moment, not publicly available \citep[e.g.][]{2014arXiv1406.7242G}}, taking into account the variations of the PSF across the field-of-view of the instrument, {\color{black} studying other possible  analysis dictionaries}, reconstructing spectral images and testing performances on poorly calibrated data. Tests of MORESANE on real observations, that will be the object of a separate paper, are on-going and promising. The results of this paper have been obtained by using MORESANE in its original version written in MATLAB. PyMORESANE, a recently developed Python implementation of MORESANE  \footnote{{The implementation is dependent only on the most common Python modules, in particular SciPy, NumPy and PyFITS.}}, is now freely available to the community under the GPL2 license (\url{https://github.com/ratt-ru/PyMORESANE)}. PyMORESANE is a complete self-contained tool which includes GPU (CUDA) acceleration and can be used on large datasets, within comparable execution time to the standard image reconstruction tools.


 \begin{figure*}[hp]
 
 \centering
 \subfloat[]{\label{halo:edge-a}\includegraphics[width=0.24\hsize]{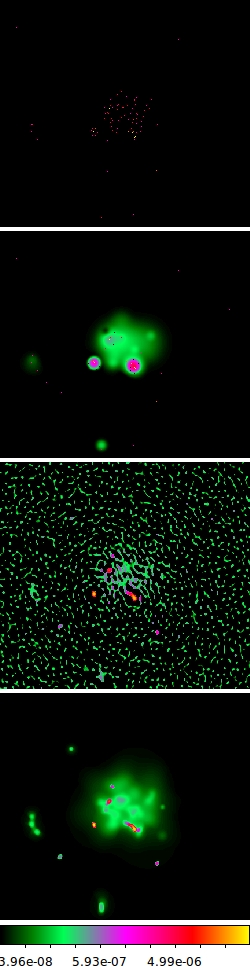}}
 \hspace{1pt}
  \subfloat[]{\label{halo:edge-b}\includegraphics[width=0.24\hsize]{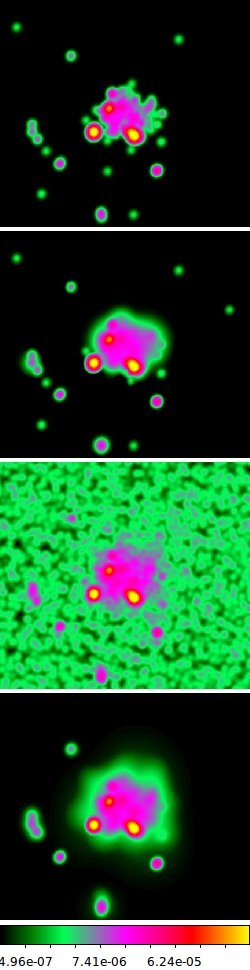}}
 \hspace{1pt}
  \subfloat[]{\label{halo:edge-c}\includegraphics[width=0.24\hsize]{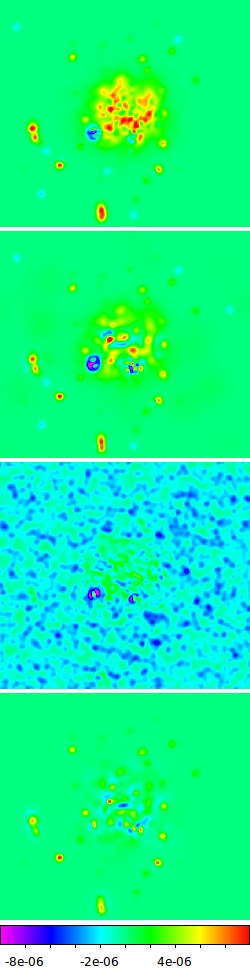}}
  \hspace{1pt}
  \subfloat[]{\label{halo:edge-d}\includegraphics[width=0.24\hsize]{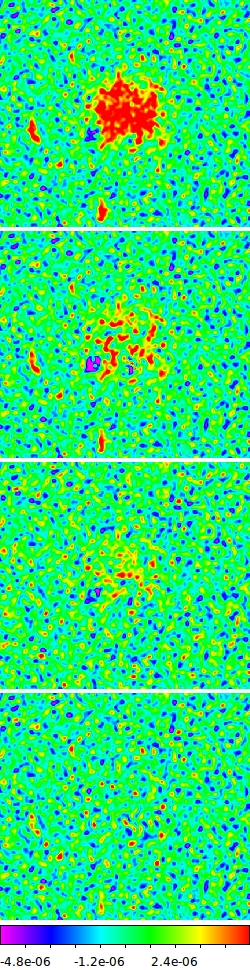}}
             \caption{ {\color{black} Reconstructed images of the galaxy cluster observations simulated with MeerKAT. The results are shown from top to bottom for H\"ogbom CLEAN, Multi-scale CLEAN, IUWT-based CS and MORESANE. From left to right, model images (\ref{halo:edge-a}),  beamed images (\ref{halo:edge-b}), error images of the beamed models with respect to the beamed true sky (\ref{halo:edge-c}) and residual images (\ref{halo:edge-d}).}
        }

       \label{halo_rslts}              
   \end{figure*}

 \begin{figure*}[hp]
\centering
 \subfloat[]{\label{M31:edge-a}\includegraphics[width=0.24\hsize, height=18.5cm]{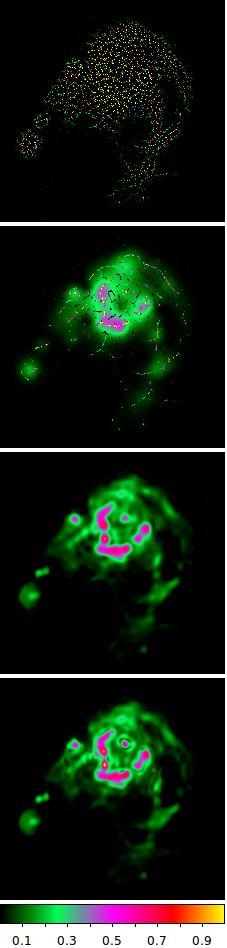}}
 \hspace{1pt}
  \subfloat[]{\label{M31:edge-b}\includegraphics[width=0.24\hsize, height=18.5cm]{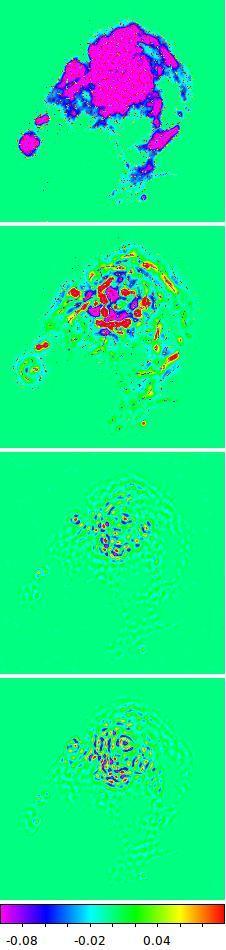}}
 \hspace{1pt}
  \subfloat[]{\label{M31:edge-c}\includegraphics[width=0.24\hsize, height=18.5cm]{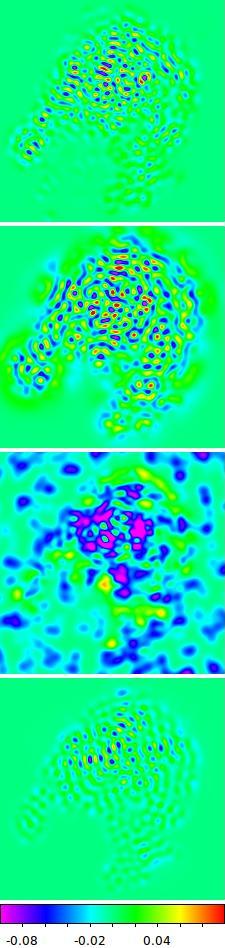}}
  \hspace{1pt}
  \subfloat[]{\label{M31:edge-d}\includegraphics[width=0.24\hsize, height=18.5cm]{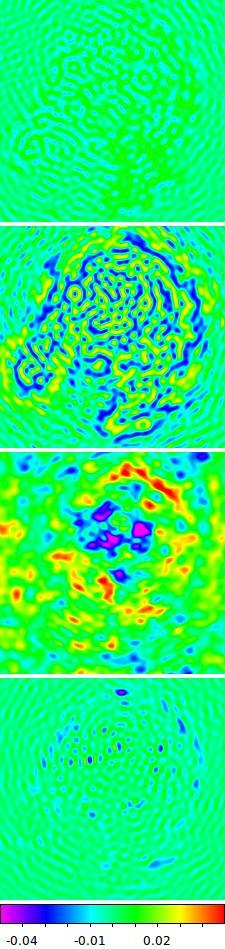}}
        
             \caption{{\color{black}Reconstructed images of M31 observations simulated with MeerKAT. The results are shown from top to bottom for H\"ogbom CLEAN, Multi-scale CLEAN, IUWT-based CS and MORESANE. From left to right, model images (\ref{M31:edge-a}),  error images of the model images with respect to the input image (\ref{M31:edge-b}), error images of the beamed models with respect to the beamed sky image (\ref{M31:edge-c}) and residual images (\ref{M31:edge-d}).} }

               \label{M31}
\end{figure*}
\begin{figure*}[hp]
\centering
\includegraphics[width=1\hsize]{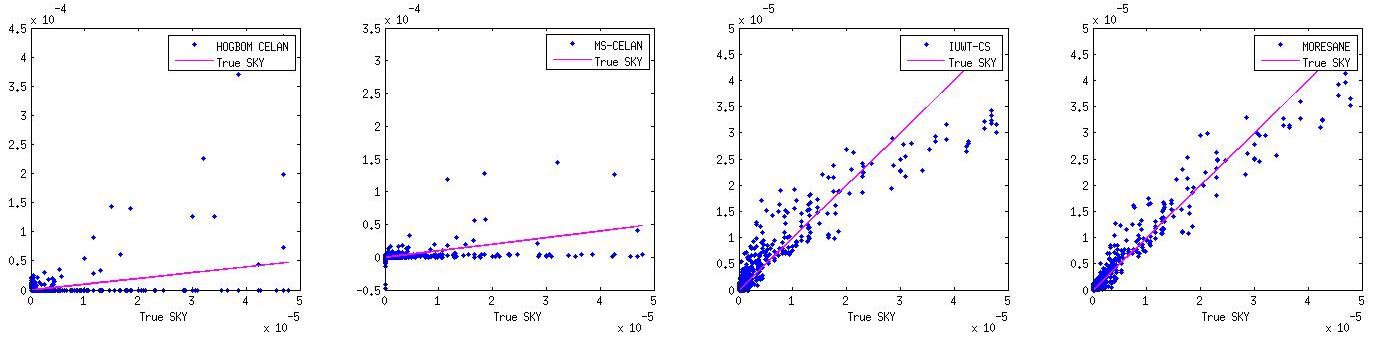}
\caption{\small{ {\color{black}From left to right: results results of the galaxy cluster recovery of H\"ogbom CLEAN, MS-CLEAN, IUWT-based CS and MORESANE. Plots of the model images ($y$-axis) against the input sky image ($x$-axis). }}}
 \label{HALOR}
 \end{figure*}
\begin{figure*}[hp]
\centering
\includegraphics[width=1\hsize]{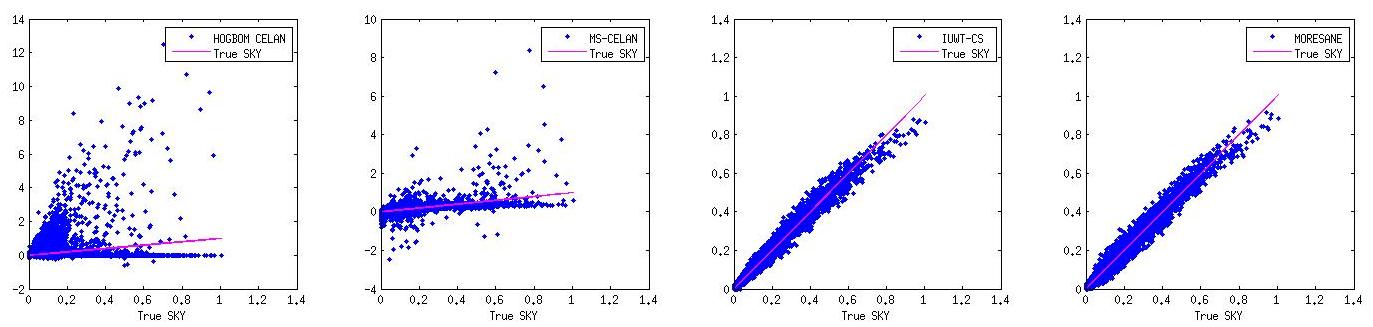}
\caption{\small{{\color{black} From left to right: results results of M31 recovery of H\"ogbom CLEAN, MS-CLEAN, IUWT-based CS and MORESANE. Plots of the model images ($y$-axis) against the input sky image ($x$-axis). }}}
  \label{M31R}
 \end{figure*}
\begin{figure*}[htp]
 
 \centering
 \subfloat[]{\label{haloS:edge-a}\includegraphics[width=0.24\hsize]{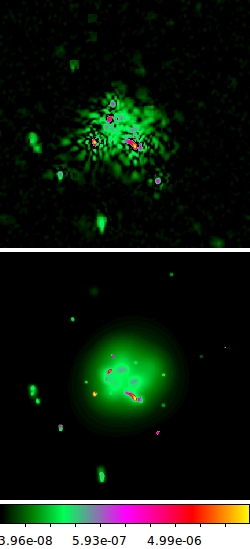}}
 \hspace{1pt}
  \subfloat[]{\label{haloS:edge-b}\includegraphics[width=0.24\hsize]{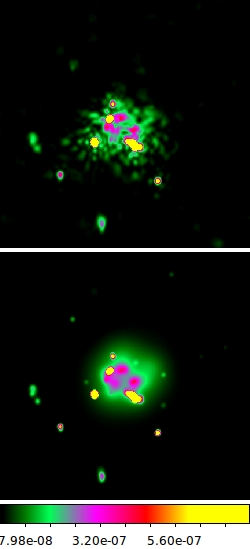}}
 \hspace{1pt}
  \subfloat[]{\label{haloS:edge-c}\includegraphics[width=0.24\hsize]{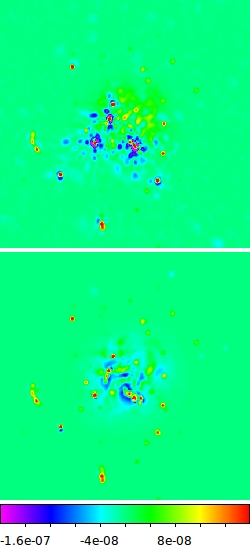}}
  \hspace{1pt}
  \subfloat[]{\label{haloS:edge-d}\includegraphics[width=0.24\hsize]{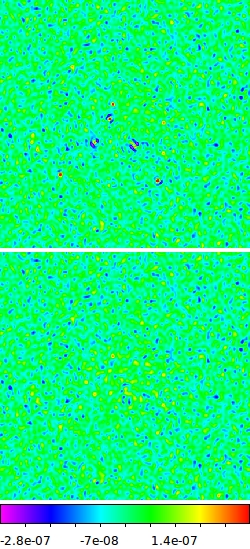}}
             \caption{{\color{black} Reconstructed images of the galaxy cluster toy simulations. The results are shown for SARA (top) and  MORESANE (bottom). From left to right, model images (\ref{haloS:edge-a}),  beamed images (\ref{haloS:edge-b}), error images of the beamed models with respect to the beamed true sky (\ref{haloS:edge-c}) and residual images (\ref{haloS:edge-d}).}
        }

       \label{haloS}              
   \end{figure*}

 \begin{figure*}[ht]
\centering
\subfloat[]{\label{SM31:edge-a}\includegraphics[width=0.24\hsize]{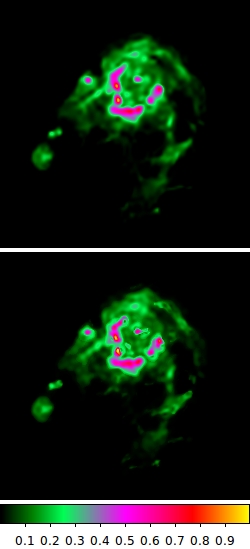}}
\hspace{1pt}
\subfloat[]{\label{SM31:edge-b}\includegraphics[width=0.24\hsize]{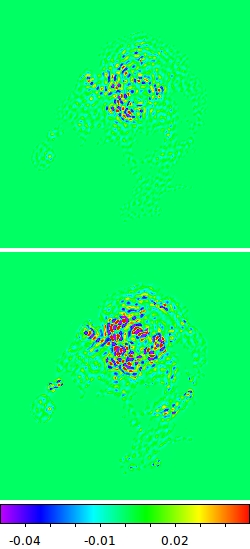}}
\hspace{1pt}
 \subfloat[]{\label{SM31:edge-c}\includegraphics[width=0.24\hsize]{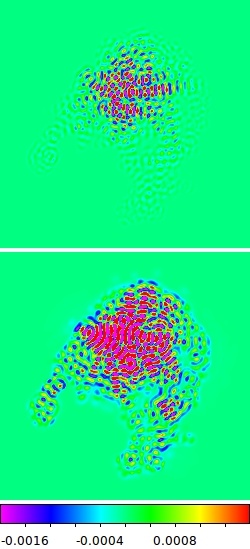}}
\hspace{1pt}
\subfloat[]{\label{SM31:edge-d}\includegraphics[width=0.24\hsize]{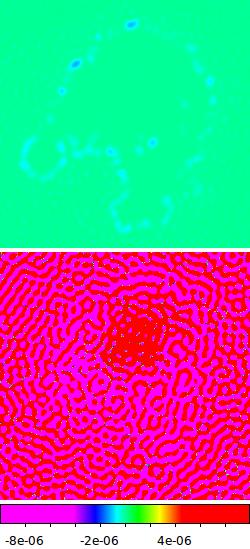}}
        
             \caption{{\color{black}Reconstructed images of M31 toy simulations. The results are shown for SARA (top) and MORESANE (bottom). From left to right, model images (\ref{SM31:edge-a}),  error images of the model images with respect to the input image (\ref{SM31:edge-b}), error images of the beamed models with respect to the beamed sky image (\ref{SM31:edge-c}) and residual images (\ref{SM31:edge-d}).} }

               \label{SM31}
\end{figure*}
 \begin{figure*}[ht]
 \centering
\includegraphics[width=0.7\hsize]{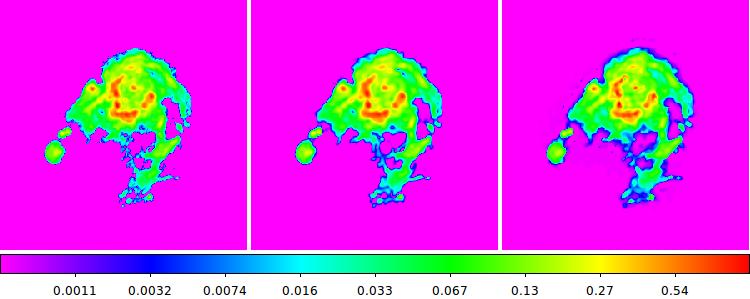}
 \includegraphics[width=0.7\hsize]{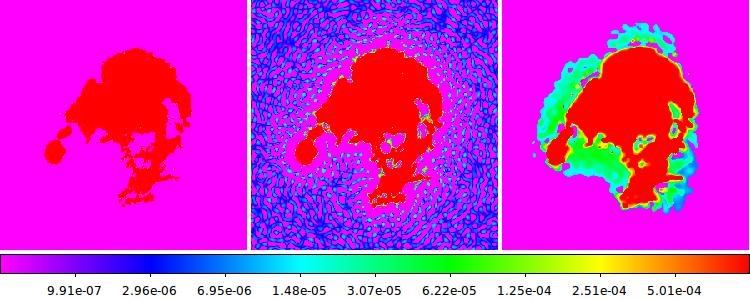}       
             \caption{{\color{black} From left to right, input model image of M31, reconstructed models respectively by SARA and MORESANE in log scale. The figures in the top and bottom lines show exactly the same images, but with different flux contrasts to highlights features within the source and its background respectively.} }

               \label{SM31log}
\end{figure*}

\begin{acknowledgements}
We warmly thank Huib Intema for very helpful comments, and Federica Govoni and Matteo Murgia for providing the galaxy cluster simulated radio map analyzed in this paper. We acknowledge financial support by the ``{\it Agence Nationale de la Recherche}'' through grants ANR-09-JCJC-0001-01 Opales and  ANR-14-CE23-0004-01 Magellan, the ``PHC PROTEA programme'' (2013) through grant 29732YK, the joint doctoral program ``{\it r\'egion PACA-OCA}'' and Thales Alenia Space (2011).
\end{acknowledgements}
\bibliographystyle{aa}

\end{document}